\documentclass[sigconf]{acmart}

\definecolor{brown}{rgb}{0.59, 0.29, 0.0}
\definecolor{darkgray}{rgb}{0.59, 0.59, 0.59}
\definecolor{tablegray}{gray}{.9}

\newcommand{\system}{MineXR}

\usepackage[utf8]{inputenc}
\usepackage{colortbl}

\usepackage{color}
\usepackage{soul}

\usepackage{xspace}
\usepackage{enumitem}
\usepackage{mathtools}

\usepackage{amssymb}
\usepackage{pifont}

\newcommand{\customtilde}{{\raise.17ex\hbox{$\scriptstyle\sim$}}}

\newcommand{\etal}{et~al.\xspace}
\newcommand{\eg}{e.\,g.,\xspace}
\newcommand{\ie}{i.\,e.,\xspace}
\newcommand{\cf}{cf.\xspace}

\newcommand{\fullwidthfigure}{\textwidth}
\newcommand{\halfwidthfigure}{.46\textwidth}

\AtBeginDocument{%
  }

\copyrightyear{2024}
\acmYear{2024}
\setcopyright{rightsretained}
\acmConference[CHI '24]{Proceedings of the CHI Conference on Human Factors in Computing Systems}{May 11--16, 2024}{Honolulu, HI, USA}
\acmBooktitle{Proceedings of the CHI Conference on Human Factors in Computing Systems (CHI '24), May 11--16, 2024, Honolulu, HI, USA}
\acmDOI{10.1145/3613904.3642394}
\acmISBN{979-8-4007-0330-0/24/05}

\begin{document}

\title[MineXR: Mining Personalized Extended Reality Interfaces]{MineXR: Mining Personalized Extended Reality Interfaces}

\author{Hyunsung Cho}
\affiliation{
  \institution{Carnegie Mellon University}
  \city{Pittsburgh}
  \state{PA}
  \country{USA}}
\email{hyunsung@cs.cmu.edu}
\orcid{0000-0002-4521-2766}

\author{Yukang Yan}
\affiliation{%
  \institution{Carnegie Mellon University}
  \city{Pittsburgh}
  \state{PA}
  \country{USA}}
\email{yanyukanglwy@gmail.com}
\orcid{0000-0001-7515-3755}

\author{Kashyap Todi}
\affiliation{%
  \institution{Reality Labs Research, Meta Inc.} 
  \city{Redmond}
  \state{WA}
  \country{USA}}
\email{kashyap.todi@gmail.com}
\orcid{0000-0002-6174-2089}

\author{Mark Parent}
\affiliation{%
  \institution{Reality Labs Research, Meta Inc.} 
  \city{Toronto}
  \state{ON}
  \country{USA}}
\email{mrkprnt@meta.com}
\orcid{0009-0004-2361-3245}

\author{Missie Smith}
\affiliation{%
  \institution{Reality Labs Research, Meta Inc.} 
  \city{Redmond}
  \state{WA}
  \country{USA}}
\email{missie.smith@gmail.com}
\orcid{0000-0001-9594-7640}

\author{Tanya Jonker}
\affiliation{%
  \institution{Reality Labs Research, Meta Inc.} 
  \city{Redmond}
  \state{WA}
  \country{USA}}
\email{tanya.jonker@meta.com}
\orcid{0000-0001-8646-5076}

\author{Hrvoje Benko}
\affiliation{%
  \institution{Reality Labs Research, Meta Inc.} 
  \city{Redmond}
  \state{WA}
  \country{USA}}
\email{benko@meta.com}
\orcid{0000-0002-2059-3558}

\author{David Lindlbauer}
\affiliation{%
  \institution{Carnegie Mellon University}
  \city{Pittsburgh}
  \state{PA}
  \country{USA}}
\email{davidlindlbauer@cmu.edu}
\orcid{0000-0002-0809-9696}

\renewcommand{\shortauthors}{Cho et al.}

\begin{abstract}
Extended Reality (XR) interfaces offer engaging user experiences, but their effective design requires a nuanced understanding of user behavior and preferences. This knowledge is challenging to obtain without the widespread adoption of XR devices. We introduce ~\system{}, a design mining workflow and data analysis platform for collecting and analyzing personalized XR user interaction and experience data. ~\system{} enables elicitation of personalized interfaces from participants of a data collection: for any particular context, participants create interface elements using application screenshots from their own smartphone, place them in the environment, and simultaneously preview the resulting XR layout on a headset. Using ~\system{}, we contribute a dataset of personalized XR interfaces collected from 31 participants, consisting of 695 XR widgets created from 178 unique applications. We provide insights for XR widget functionalities, categories, clusters, UI element types, and placement. Our open-source tools and data support researchers and designers in developing future XR interfaces.
\end{abstract}


\begin{teaserfigure}
  \includegraphics[width=\textwidth]{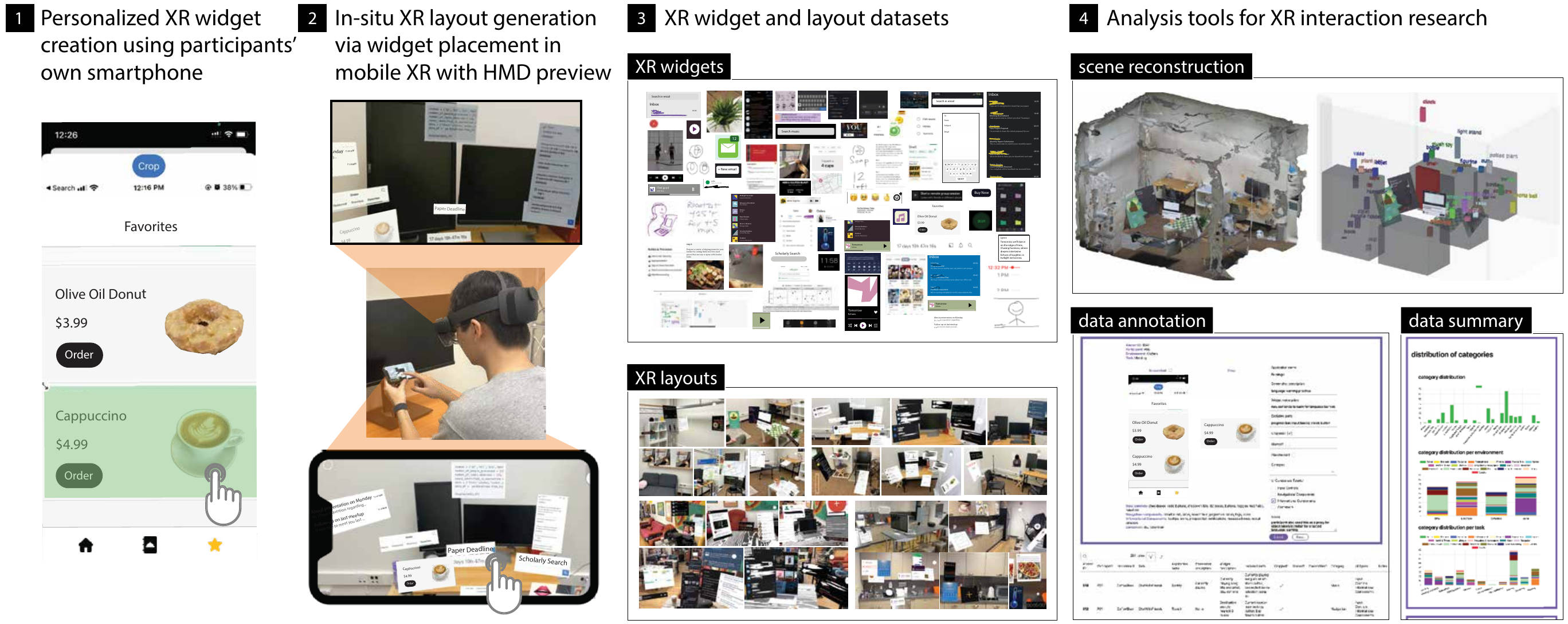}
  \caption{\system{}'s data mining workflow enables researchers to collect and analyze users' personalized XR interfaces in various contexts. We release our \system{} dataset, consisting of 695 XR widgets in 109 unique XR layouts.}
  \label{fig:teaser}
  \vspace{1em}
\end{teaserfigure}


\maketitle

\section{Introduction}
Extended Reality (XR) has the potential to transform how users interact with digital content and user interfaces.
Within the context of everyday always-on XR, we envision users wearing next-generation head-mounted displays (\eg XR glasses or contact lenses) that enable ubiquitous access to digital information, complementing or replacing their current mobile devices such as smartphones and smartwatches.
This will enable users to unobtrusively and ubiquitously access information for tasks such as navigation, messaging, cooking, or web browsing.

Research in XR has explored a wide range of factors for improving user experience and performance, from advanced interaction techniques~\cite{Nuernberger_Ofek_Benko_Wilson_2016} and representation of content~\cite{Lee2022, Lindlbauer_ContextAwareMR_2019, julier2000information, diverdi2004level, tatzgern2016adaptive, ghouaiel2014adaptive}, to view management~\cite{huynh2022layerable, Lu_Davari_Lisle_Li_Bowman_2020, bell2001view, azuma2003evaluating, grasset2012image} and adaptive layout systems~\cite{Lindlbauer_ContextAwareMR_2019,Cheng_SemanticAdapt_2021}.
It is still unclear, however, \textit{what} virtual content users would find most useful, or like to see and interact with, in everyday XR. 
Specifically, we lack an understanding of what kind of applications and digital functionalities users will require, and how contextual aspects such as activity and environment impact this choice.
In addition, it is still undetermined whether an application-based interface approach that dominates current desktops and smartphones is desirable for XR or whether a more fine-grained functionality-based approach based on individual widgets is preferred by users.
The former encapsulates functionality in a clear, hierarchical manner, but might not make use of flexible presentation opportunities in XR, where interfaces can be placed nearly anywhere with arbitrary amounts of information.
The latter allows flexible presentation but might lead to XR layouts that are perceived as disconnected or tedious.

To address this challenge, we contribute the (1) \system{} data collection and analysis platform and (2) \system{} dataset of personalized XR layouts.
The data collection tool (Figure~\ref{fig:teaser}) provides a novel workflow that enables researchers to collect custom XR layouts in situ using personal digital content. 
Participants of a data collection use their personal smartphones to extract content, specifically screenshots of their apps and frequently used websites in different environments and for various activities.
Leveraging our platform, participants create fully personalized XR widgets and compose XR layouts freely in space using their smartphone. 
They concurrently view the layout through an optical see-through head-mounted display.
We refer to \textit{widgets} as individual or compound virtual UI elements containing specific functionalities of an application, which participants place in the XR environment, \eg a play/pause button for music (individual) or a lyrics+play control panel (compound) from a music-playing app. Other examples of widgets include real-time bus location tracker, favorite coffee order, today's schedule, etc. 
Researchers can review and annotate the collected data in the web-based \system{} data exploration and annotation tool and a 3D layout viewer.

Leveraging our platform, we contribute a dataset of personalized XR layouts to answer questions regarding which functionalities are desirable in what context, and how widgets are composed.
The \system{} dataset consists of 109 XR layouts comprising of a total of 695 widgets, which were created by 31 participants for activities in everyday lives (\eg focus work, meeting with colleagues, relaxing, cooking, and making coffee), located in four different real-world environments (office, living room, kitchen, coffee shop).
We refer to a combination of activity and environment as a \textit{scenario} (\eg \textit{working} in \textit{office}, \textit{relaxing} in \textit{living room}).
Our dataset is annotated with information about what widgets and functionalities were extracted from screenshots for XR, what categories of applications were used (\eg productivity, entertainment), which widgets were clustered in space, and what types of UI elements were used (informational component, navigational component, input control).
We connect this data with information about activity and environment to gather more detailed insights on XR layouts in different contexts.
We present analysis results focused on the categories, functionalities, clusters, and UI elements of the collected data. 
We contribute a set of design recommendations for the creation of XR interfaces based on the analysis.
For example, future XR interfaces should provide \textit{functionality}-based XR interfaces beyond the application-based paradigm.

Our dataset complements current approaches for gathering data in XR scenarios that rely on presenting users with predetermined applications~\cite{Lu_Davari_Lisle_Li_Bowman_2020, Davari_Lu_Bowman_2022}.
As participants created individualized XR layouts using their personal digital content, our collected data better reflects their realistic preferences and requirements for layouts.
The dataset enables researchers to draw insights into content selection, layout consideration, and context-aware usage of future XR applications, beyond what is presented in this paper.
Furthermore, we believe it can be a valuable resource for developing novel computational approaches for XR UIs that address challenging problems such as context-aware functionality recommendation and assistive placement.
We believe that our dataset is a step towards a future where everyday XR interfaces become feasible and beneficial for users, without requiring researchers to wait for the release of future headsets that can be worn in an always-on manner.
Our dataset, data collection workflow, and analysis tools are available at \textit{\url{https://augmented-perception.org/publications/2024-minexr.html}}\footnote{Some figures in the paper include mockups to de-identify the original application, while retaining essential design elements and functionalities of the original screenshots and widgets.}.
\section{Related Work}
MineXR enables interactive mining and analysis of personalized XR interfaces using existing UIs.
Insights drawn from data collected with MineXR can be used to develop adaptive user interfaces for XR.  
Here, we discuss prior literature on design and adaptation of XR user interfaces as well as the use of existing UIs to inform design.
\vspace{-0.5em}
\subsection{Adaptive User Interfaces for XR} 
\label{sec:related_work_adaptive_UI}
Existing approaches to building adaptive user interfaces for XR focus on the presentation of virtual elements, \ie \textit{where} and \textit{how} to place virtual elements in various environments.

As summarized by Qian~\etal\cite{Qian_He_Hu_Wang_Ipsita_Ramani_2022}, there exists a wide variety of approaches to anchoring XR contents in the virtual world, from tracking-based approaches~~\cite{spatial, kwan2019mobi3dsketch} to marker-based approaches~\cite{Seichter_Looser_Billinghurst_2008, Kato_MarkerTracking_1999, jo2016ariot, wagner2007artoolkitplus, huo2018scenariot, romero2018speeded} that anchor XR contents using fiducial markers or specific objects.
Geometry-based and semantic-based approaches employ more adaptive strategies that are based on the understanding of a target scene. 
Geometry-based approaches align XR contents with edges, planes, and meshes~\cite{Nuernberger_Ofek_Benko_Wilson_2016, Ens_Ofek_Bruce_Irani_2015, gal2014flare, herr2018immersive, du2020depthlab, fender2018optispace, han2023blendmr}. 
Semantic-based approaches leverage the semantic relationships between the physical entities in the scene and virtual elements~\cite{Cheng_SemanticAdapt_2021, tahara2020retargetable, dong2021tailored, han2020live, lang2019virtual, liang2021scene}, or the current context of the user~\cite{fender2017heatspace}, including physical activity, \eg walking \cite{Lages_Bowman_2019c} and cognitive load~\cite{Lindlbauer_ContextAwareMR_2019}.

Another line of research investigates the interaction behaviors with different layouts of XR user interfaces.
Lu~\etal\cite{Lu_Davari_Lisle_Li_Bowman_2020, Lu_Bowman_2021} and Davari~\etal\cite{Davari_Lu_Bowman_2022} evaluated the Glanceable AR paradigm for accessing information in AR HMDs, in which secondary information resides at the periphery of vision, staying unobtrusive yet accessible through a quick glance.   

All of these approaches consider where to place \textit{pre-defined} virtual elements in the real-world scene and how to position them in an adaptive layout.
Despite the efforts on adaptive presentation of virtual elements, it is not clear what \textit{content} should be presented in XR interfaces.
Specifically, it is unclear what kinds of virtual elements users would want to use in XR for everyday uses, and how much of the content users would like to see in an XR environment at a time.
The lack of knowledge is partly due to the yet-immature XR devices for everyday uses. 
However, this is a chicken-and-egg problem, where understanding the potential usability can provide insights and motivation to develop the XR devices further to support these use cases.
The \system{} platform aims to alleviate this problem by enabling researchers to collect data on these aspects of XR content creation.
%
\vspace{-1em}
\subsection{Leveraging Existing UIs for Design}
In website and mobile app designs, data-driven models have been introduced to help designers understand current practices and trends through example designs~\cite{Deka_RicoMobile_2017, Kumar_Webzeitgeist_2013, Kumar_Bricolage_2011}.
In web design, Kumar~\etal introduced an example-based web design retargeting model~\cite{Kumar_Bricolage_2011} and a design mining technique~\cite{Kumar_Webzeitgeist_2013} for a large-scale collection of web designs that is used for understanding design demographics, design curation, and supporting data-driven design tools. 

In mobile app design, Deka~\etal collected a repository of user flows~\cite{Deka_ERICA_2016} and mobile app designs~\cite{Deka_RicoMobile_2017} through a combination of crowdsourcing and automated design and interaction mining from Android apps. 
This large-scale mobile app design dataset supports design search, UI layout generation, UI code generation, user interaction modeling, and user perception prediction. 
Annotated datasets of mobile UI screens can serve as a basis for generating accessibility metadata labels for UI components~\cite{zhang2021screen}, generating screen summaries~\cite{Wang_Screen2WordsAutomatic_2021}, and performance testing~\cite{Deka_ZIPTZeroIntegration_2017}.
Additionally, these datasets can be leveraged for semantic embeddings~\cite{Li_Popowski_Mitchell_Myers_2021,Bai_Zang_Xu_Sunkara_Rastogi_Chen_Arcas_2021} in mobile UI to decode into other forms of UIs such as automatic UI generation~\cite{Fischer_Campagna_Xu_Lam_2018,gajos2004supple}, smartphone shortcuts~\cite{Li_Azaria_Myers_2017}, custom UI mashup~\cite{Lee_Kim_Kim_Lee_Kim_Song_Ko_Oh_Shin_2022,kim2019x}, and conversational UIs~\cite{Li_Riva_2018, Wang_EnablingConversational_2023}.

Unlike website and mobile app designs, such data-driven model approaches in XR for everyday uses cannot be directly adopted primarily because XR as a platform is not yet as prevalent as web or mobile apps and thus has a small user base. 
More importantly, XR user interfaces are not constrained inside a screen, exponentially enlarging the possible interaction space, contexts, and settings, which complicates the collection of interface designs with significant coverage of the design space. 
The in-situ interactive mining workflow of \system{} enables researchers to elicit contextual, personalized XR interfaces through in-situ creation of XR widgets and 3D spatial placement in diverse real-world contexts.

\subsection{Prototyping for XR Interfaces}
Most research on XR interfaces has focused on enhancing designers' prototyping and authoring process when designing specific XR experiences.
%
These designer tools support the creation of 3D AR content or rapid prototyping of interactive AR experiences ~\cite{macintyre2004dart, Guven_Authoring3D_2003, Guven_MobileAugmented_2006, Gasques_WhatYou_2019, Nebeling_ProtoARRapid_2018}. 
For example, Pronto~\cite{Leiva_ProntoRapid_2020a} and other works by Leiva~\etal\cite{Leiva_EnactReducing_2019, Leiva_RapidoPrototyping_2021} enable rapid prototyping for XR interfaces using sketches and enaction. They ground their design based on the recurring issues in XR design they identified: creating and positioning 3D assets, handling the changing user position, and orchestrating multiple animations, which echo the major challenges identified by MacIntyre~\etal\cite{macintyre2004dart}.
These prototyping and authoring tools assist XR experience designers in imagining their creations in action as a process of design iteration. 

Researchers have also explored in-situ and ad-hoc authoring experiences through immersive authoring \cite{Huo_windowshaping_2017, Wang_capturar_2020, Yue_scenectrl_2017}.
VR has been a popular platform to support immersive authoring~\cite{Mine_ISAAC_1995, Zhang_FlowMatic_2020} with the advantage of allowing designers to explore scenarios free from temporal and spatial limitations
\cite{Jetter_InVREverything_2020,Xia_Spacetime_2018}.
To remedy the hardships in conducting XR prototyping in the real world, creating experiences in VR based on the virtual replica of the physical environment has been also widely adopted~\cite{Wang_DistanciAR_2021, Ens_Ivy_2017, Prouzeau_CorsicanTwin_2020}.
For example, Qian ~\etal proposed an integrated workflow, ScalAR~\cite{Qian_He_Hu_Wang_Ipsita_Ramani_2022}, which provides a VR authoring environment where a designer can create semantic-level associations of the XR contents in each virtual scene synthesized from real-world sample scenes. 
Its model learns the semantic associations from the designer's demonstrations and adaptively renders AR content in different target environments.

These works have a different goal of streamlining the prototyping and authoring process of XR interface creation through VR with a focus on \textit{where} and \textit{how} a given XR content would be presented.
Our work, on the other hand, focuses on \textit{what content} will be used in XR and \textit{in which format} the content will be presented.
We achieve this through a bottom-up approach to take advantage of the rich array of personalized interfaces in mobile applications and websites.






\begin{figure*}[t]
    \centering
    \includegraphics[width=\fullwidthfigure]{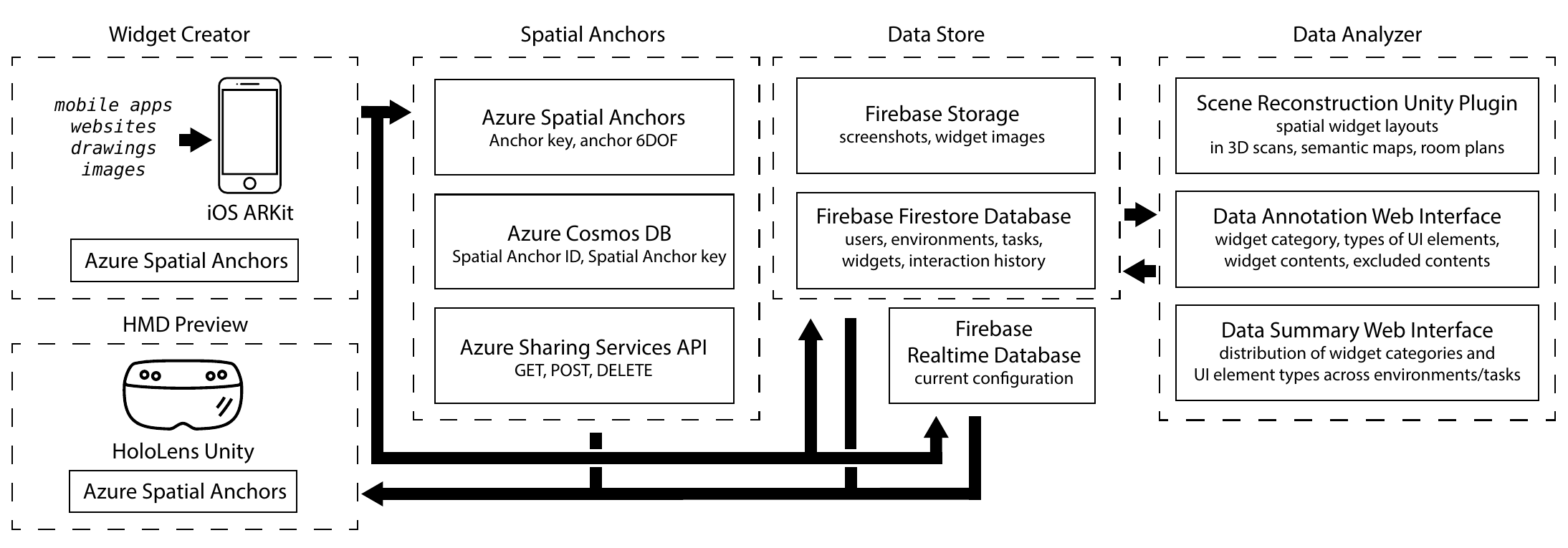}
    \vspace{-2em}
    \caption{Architecture of the \system{} data collection and analysis platform.}
    \label{fig:architecture}
\end{figure*}

\section{\system{}} \label{sec:system}
We present \system{}, a novel workflow that enables researchers to collect, or `mine', rich data on personalized XR layouts for gaining insights into future usage of XR and for adaptive XR systems. 
\vspace{-0.5em}
\subsection{Design Principles}
We took inspiration from XR authoring tools~\cite{Nebeling2018_trouble}, adaptive interfaces \cite{Cheng_SemanticAdapt_2021,Belo_Lystb_2022}, cross-device interaction (\cf Brudy~\etal\cite{Brudy2019cross_device}), and prototyping toolkits~\cite{herskovitz2022xspace, nebeling2020mrat}.
Our platform is designed to enable researchers to collect rich data for future everyday XR experiences.
This, in turn, means that we aim to provide a fast, flexible, and well-suited experience for participants who create the XR layouts. 
Our design goals are targeted toward participants of a data collection.
\subsubsection*{Personalized Prototyping} 
Using predefined widgets introduces designer bias and can limit our understanding of what content and UIs may be desirable in the future.
To circumvent this issue, we use a \textit{bottom-up approach} by taking advantage of the rich app ecosystem available on smartphones, inspired by work on designing personalized websites~\cite{Kumar_Bricolage_2011}.
Specifically, participants create their personalized XR UIs from apps and content available within their own device.
%
\subsubsection*{In-situ Prototyping} 
Typically, prototyping is restricted to lab settings, which limits participants' abilities to contextualize and understand their varying needs.
We enable in-situ prototyping, where participants are immersed in diverse real-world scenarios and environments, to enable rich and individualized data collection and understanding.  
We develop an untethered setup, with a smartphone and HMD, connected via cloud infrastructure, and use spatial anchors, synchronized between the two devices, so that participants can create XR layouts in near-arbitrary uninstrumented locations.
%
\subsubsection*{Decomposing Apps to Functionalities} 
We believe that personalized experiences require a fine-grained selection of \textit{functionalities} rather than just the selection of apps, similar to work on personalized UI mashup~\cite{Lee_Kim_Kim_Lee_Kim_Song_Ko_Oh_Shin_2022,kim2019x}.
We enable flexible composition of widgets into XR layouts through cropping and free-form mockups.
%
\subsubsection*{Zero Programming}
We aim to enable participants with no prior experience in programming or interface authoring to create personalized XR layouts.
There exists a rich set of tools for content creation and layout adaptation in XR.
To the best of our knowledge, however, no previous approach has enabled participants to create personalized interfaces based on their own existing digital content (\ie apps).
We thus adopt well-known paradigms of smartphone interactions (\eg built-in functionalities for screenshots or image annotation) and use advanced XR interactions only for display purposes.
This simplifies the creation of XR layouts drastically.
\begin{figure*}[t]
    \centering
    \includegraphics[width=\fullwidthfigure]{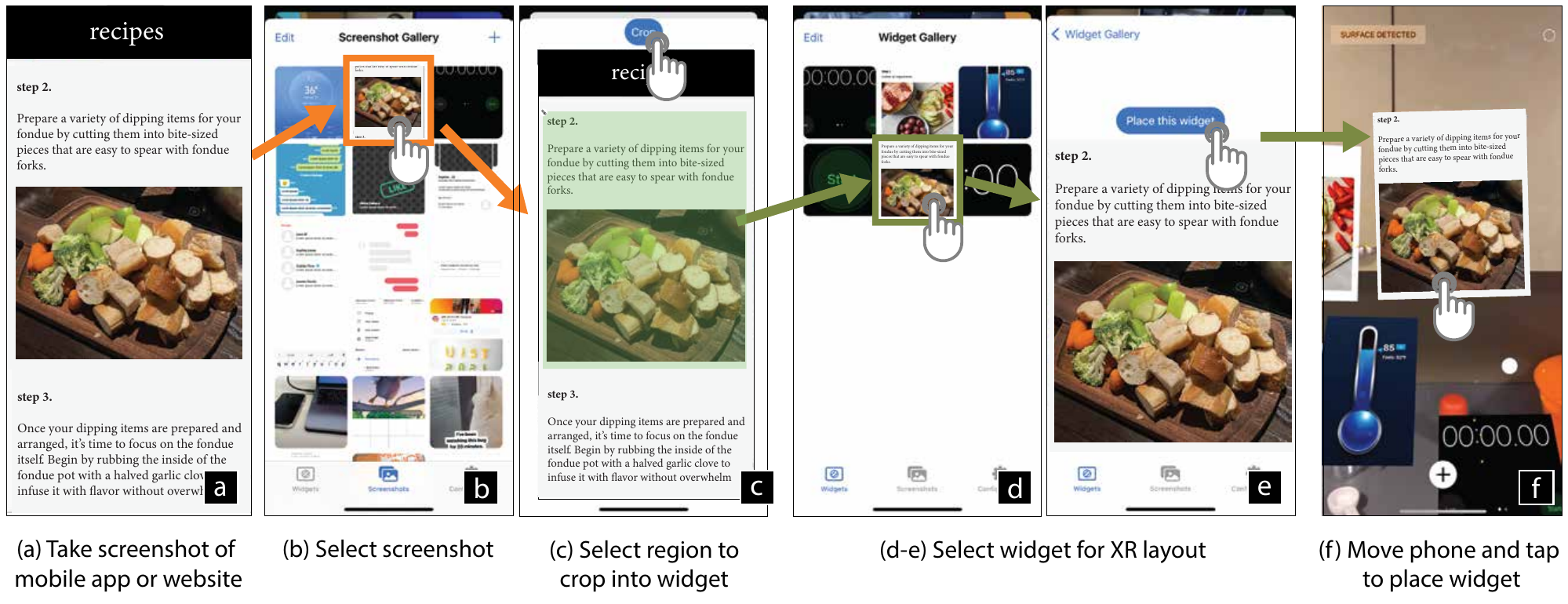}
    \caption{Widget creation and placement workflow. \system{} provides a smartphone-based widget creation and XR placement workflow. A participant can create new widgets from screenshots of mobile apps or websites. Then, they can create XR widget layouts by moving the phone to the desired widget location and tapping on the screen to place.}
    \label{fig:widget_collector}
\end{figure*}

\subsection{Overview of Data Collection}
We explain the workflow, assuming that researchers have set up the \system{} data collection platform, and guided participants to a real-world environment or lab.
In a given environment and task, participants first take screenshots of mobile applications and websites, create sketched mockups, or choose existing media from their phone~(Figure~\ref{fig:architecture}. Widget Creator).
Using this media, participants create \textit{widgets} for XR usage.
Widgets are the individual or compound virtual UI elements containing specific functionalities of an application, which participants place in the XR environment.
For example, in Figure~\ref{fig:widget_collector}, the cropped step-by-step recipe element is a widget, which serves the functionality of showing a step-by-step recipe.

To create an XR layout, participants place each widget in the current environment by moving a smartphone running the \system{} application to the desired physical location and tapping on the screen.
\system{} continuously shows a preview of the composed XR layout in the head-mounted display (HMD)~(Figure~\ref{fig:architecture}. HMD Preview).
This facilitates the process of envisioning the actual usage of this XR layout.

\system{} consists of five integrated components: widget creator, widget placement, cloud storage, HMD preview, and data analyzer.
The widget creator, widget placement, and the HMD preview provide participants with the ability to create and view XR layouts, enabled by the \system{} cloud storage component~(Figure~\ref{fig:architecture}. Spatial Anchors and Data Store).
The data analyzer enables researchers to work with the collected data~(Figure~\ref{fig:architecture}. Data Analyzer).
Before or after data collection, researchers can create virtual replica of the data collection environments using a 3D scan and a semantic 3D map via commercial software for later use in data analysis.

\subsection{Widget Creation}
The widget creator~(Figure~\ref{fig:architecture}) enables participants to specify different widgets for everyday XR usage based on their own mobile apps, websites, photos, and drawings.
It captures \textit{what} content and interfaces participants might want to typically use in XR by collecting visual representations of desired functionalities in a given context. 

Figure~\ref{fig:widget_collector} summarizes the workflow for creating a personalized XR layout from participants' perspective. 
The creator functions in three steps.
The participant first takes screenshots of mobile apps and websites or drafts a new mockup drawing using the iOS screenshot and markup features~(Figure~\ref{fig:widget_collector}a).
They can redact any sensitive or private information from screenshots using the in-built markup functionality.
Our custom iOS app, written in Swift, enables participants to import screenshots or mockup drawings of their choice~(Figure~\ref{fig:widget_collector}b).
Additionally, \system{} provides placeholders for different apps (\eg email, messenger, dating) that can be used instead of participants' actual app.

Next, the app provides a cropping functionality where participants can crop different parts of the imported screenshots into widgets~(Figure~\ref{fig:widget_collector}c). 
The cropped widgets are stored as separate files and displayed in the widget gallery~(Figure~\ref{fig:widget_collector}d). 
All screenshots and widgets are uploaded to the \system{} cloud storage~(Figure~\ref{fig:architecture}. Data Store) along with their metadata, including information on the hierarchical relationship between screenshots and widgets.

\subsection{In-situ Widget Placement}
Once the participant selects a widget to include in their XR layout~(Figure~\ref{fig:widget_collector}d-e), they place the widget as an anchor in the XR scene through a mobile view implemented with iOS ARKit. 
Participants specify the 3D position of the widget in the scene by using the phone as a proxy of the widget.
When the participant brings the phone to the desired location for the widget and taps on the screen, the app creates a spatial anchor at the phone's position in the world coordinate, tracked by ARKit~(Figure~\ref{fig:widget_collector}f). 
The position of the last placed widget can be adjusted multiple times following the same procedure.
As an alternative, the widget placement component provides a raycasting selection method, enabling participants to select and re-place previously positioned widgets.
\begin{figure*}[t]
    \centering
    \includegraphics[width=\fullwidthfigure]{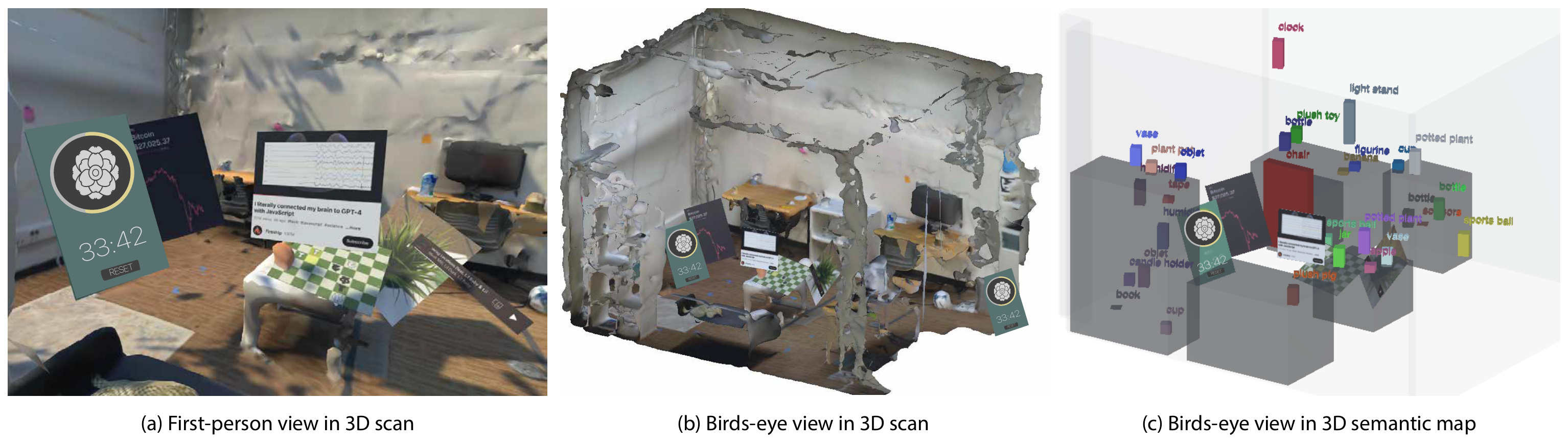}
    \caption{Example screenshots taken from \system{}'s interactive scene reconstruction Unity plugin, including a participant-generated layout. Researchers can reconstruct the spatial layout of each scenario with an optional overlay of the environment's 3D scan or semantic map. The 3D scan (a,b) in this example was generated using Polycam, the semantic map (c) using Apple RoomPlan.}
    \label{fig:scene_reconstruction}
    
\end{figure*}
\begin{figure*}[t]
    \centering
    \vspace{-1em}
    \includegraphics[width=\fullwidthfigure]{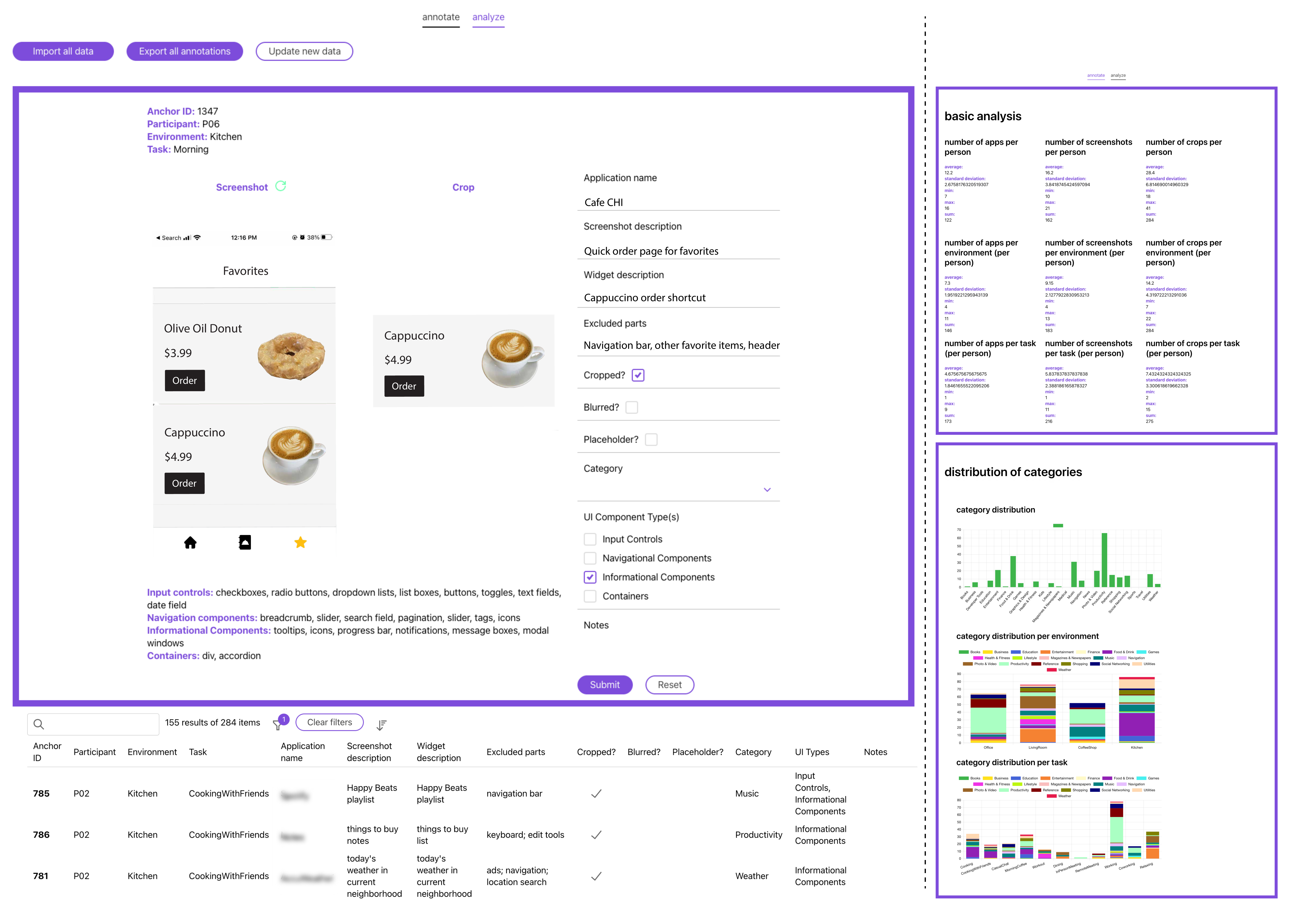}
    \caption{Web-based data annotation and summary interface. The annotation interface (\textit{left}) loads each instance data of widget placement and displays the corresponding anchor ID, participant ID, environment, task, screenshot image, and widget image of the widget.
    The interface provides a form interface including fields for data annotation, \eg application name, screenshot description, widget description, excluded parts, type of UI element(s), application category, etc.
    Below the purple annotation box is an interactive data table where researchers can browse through the annotations by search, filters, and sorts.
    The summary interface (\textit{right}) provides basic statistical summary and visualizations of the annotated data. }
    \label{fig:data_analyzer}
\end{figure*}
\subsection{Cloud Data Storage}
\system{}'s server consists of Azure Spatial Anchors (ASA) manager (\autoref{fig:architecture}. Spatial Anchors) and Firebase data store (\autoref{fig:architecture}. Data Store). 
Once the local anchor is successfully created by placing an XR widget, the iOS app creates a cloud anchor using ASA SDK\footnote{https://azure.microsoft.com/en-us/products/spatial-anchors}. 
\system{} stores the position and orientation of the created anchors. 
The information is retrieved using a client-side SDK for widget placement on the phone, and for the XR preview shown through the HMD.
On the server side, \system{} stores the anchor including metadata via ASA services that can be accessed via Azure Cosmos DB and Azure Sharing Services.

\system{} employs another data store using Google Firebase\footnote{https://firebase.google.com/} Storage and Firestore Database for screenshots, widget images and other widget-related information. 
Each widget instance stores information about its participant, environment, task, container screenshot, interaction history~(add/update), and anchor ID. 
This anchor ID is used to query the anchor's position and rotation.
The participant, environment, and task information are entered as text input in the iOS app by the experimenter at the beginning of data collection in every scenario. A scenario refers to an environment-task pair. For example, working in the office is a unique scenario from working in a coffee shop (different environment) or having a meeting in the office (different task).

\subsection{Head-mounted Display (HMD) Preview}
\system{} enables cross-platform sharing and provides participants with an XR preview through an HMD~(\autoref{fig:architecture}. HMD Preview).
The HMD application was implemented in Unity, also using the ASA SDK, and deployed to a Microsoft HoloLens 2 headset.
The Unity application continuously queries our service for widget updates.
When it detects a new widget, it creates a new object at the anchor position and rotation (obtained from ASA) and applies the texture retrieved from the widget image in Firebase Storage.
In the current implementation, the environment and task information are input as metadata in the iOS app. 
This real-time synchronization approach, however, considers potential extensions in future implementations, \eg automatically detecting a participant's dynamically changing environment and task through human activity recognition based on continuous inertial measurement unit (IMU) and audio sensing.

\subsection{Data Analyzer} \label{sec:data_analyzer}
The data analyzer~(\autoref{fig:architecture}. Data Analyzer) supports researchers by providing an interface to view and annotate the collected data.
It offers a 3D spatial layout and scene reconstruction tool in a Unity plugin~(Figure~\ref{fig:scene_reconstruction}) and a web-based data annotation and analysis interface (Figure~\ref{fig:data_analyzer}).

\system{} collects the following raw information, which is available in the dataset:
\begin{itemize}
    \setlength\itemsep{0em}
    \item Real-time participant position and rotation in world coordinates
    \item World coordinates of widgets
    \item Layout placements of widgets per scenario
    \item Screenshots and cropped widget images along with a mapping between screenshot and created widget
    \item Interaction history (add/update) per widget
\end{itemize}

On top of these raw data, researchers can provide semantic annotations of the cropped widgets for richer analysis using the data analyzer, for examples included in our dataset:
\begin{itemize}
    \setlength\itemsep{0em}
    \item App names
    \item Screenshot descriptions
    \item Widget descriptions and functionalities
    \item Widget categories
    \item UI components in the widget
    \item Widget clusters
\end{itemize}

\subsubsection{Scene Reconstruction}
\system{} includes a Unity plugin that enables researchers to reconstruct the participant-generated spatial layout (Figure~\ref{fig:scene_reconstruction}). 
The plugin takes the participant ID and environment as inputs, and retrieves all the widget anchors and their interaction history from our server to reconstruct the scene including all widgets. 
The plugin provides two options for reconstruction: showing all widgets at once and step-by-step interaction history browsing using arrow keys.
By default, all widgets and their spatial layouts can be rendered within an empty 3D space in the custom Unity plugin.
Researchers using \system{} may obtain a 3D scan and a semantic 3D map via commercial software, such as Polycam\footnote{https://poly.cam/} and Apple RoomPlan\footnote{https://developer.apple.com/augmented-reality/roomplan/}, respectively. 
If 3D scans of the environment are available, the reconstructed widgets and layouts can be viewed within this 3D space (Figure~\ref{fig:scene_reconstruction}a-b) or a semantic 3D map (Figure~\ref{fig:scene_reconstruction}c).

\subsubsection{Data Annotation and Summary}
\system{} contains an integrated web interface for data annotation and analysis, implemented in React JS, shown in Figure~\ref{fig:data_analyzer}. 
Researchers can load all widget data stored in the server, \eg screenshots, widget images, environment, and scenario information.
Researchers annotate the data using a list of form fields for each attribute, \eg text descriptions of the screenshot and cropped image, types of UI elements, and category of the widget from a list of 27 possible categories of the Apple App Store\footnote{\url{https://developer.apple.com/app-store/categories/}, retrieved March 16 2023}, such as Business, Entertainment, or Finance.
This step could be further automated in the future, \eg by leveraging information from the RICO dataset~\cite{Deka_RicoMobile_2017}.
\system{} includes an autocompletion suggestion feature to expedite the annotation process of repetitive contents.
Researchers can browse through the annotated data using an interactive data table that supports data search, filtering, and sorting.
\system{} also provides an analysis dashboard showing a real-time overview of the annotated data.
The information includes the average, minimum, and maximum number of screenshots created per participant for each environment or task, and visualizes the distribution of categories and types of UI elements too. 
Finally, researchers can annotate groups and clusters of widgets in combination with the scene reconstruction Unity plugin.
This provides valuable insights into potential combinations of functionalities for a given context, especially when viewed in the 3D scan, or in the environment where participants created the layout by leveraging our preview functionality.
We use our tool for the analysis of our dataset.
\section{\system{} Data Collection} \label{sec:data_collection}
\system{} enables researchers to collect data on what contents participants would place in an XR environment, and gather insights about layout, grouping, and desired functionalities.
In the following, we report on the design of the data collection.
Our method and later analysis serve as a blueprint for future research using \system{}.

\subsection{Scenarios and Setup}
We collected data in four separate environments.
In each environment, participants created XR layouts for two separate tasks (\eg working, chatting with friends).
A \textit{scenario} is a combination of environment and task (\eg working in kitchen, relaxing in coffee shop).
Each participant was asked to create a total of four layouts for different scenarios, \ie two environments $\times$ two tasks.
The protocol was approved by the local university institutional review board.
\subsubsection{Environments}
\begin{figure}[t]
    \centering
    \includegraphics[width=\halfwidthfigure]{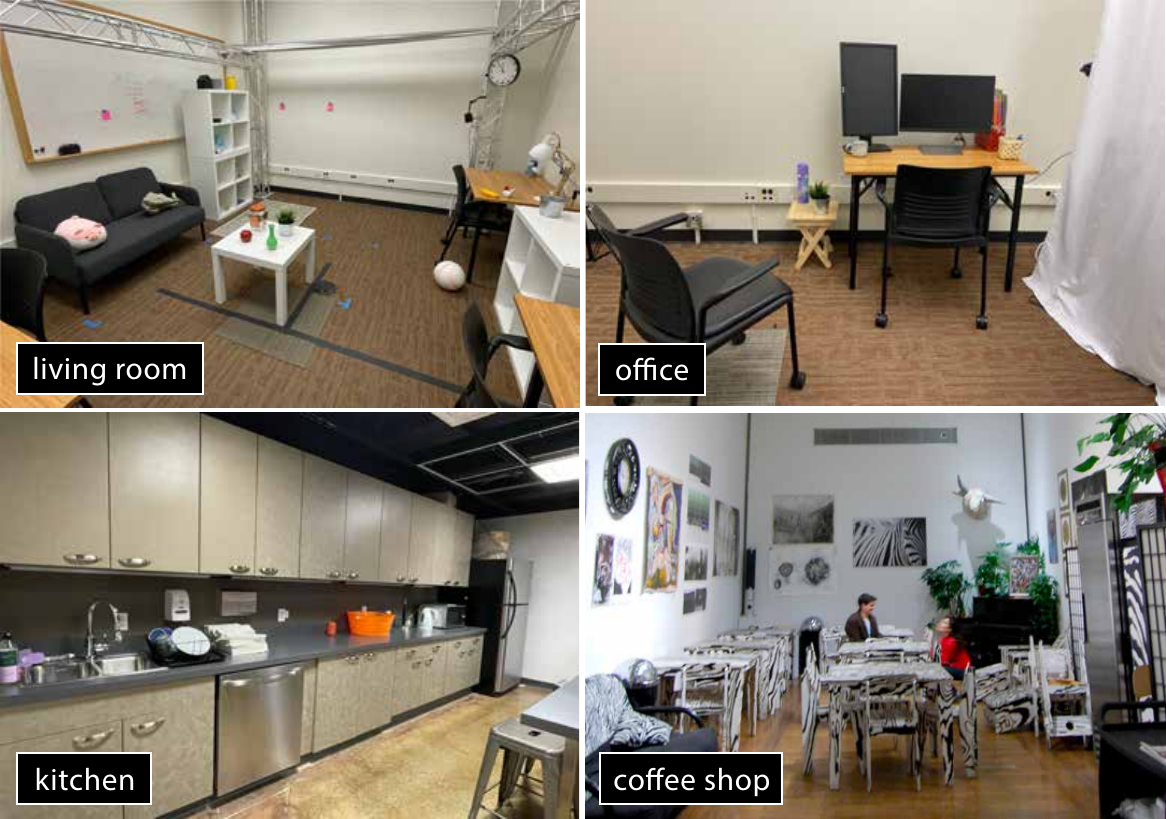}
    \caption{The four environments in which we collected data. Each participant created XR layouts in two of the four environments.}
    \vspace{-0.5em}
    \label{fig:study_environments}
\end{figure}

We collected data in an office and a living room, both set up in an experimental room.
Additionally, we collected data in a kitchen and a coffee shop.
Figure~\ref{fig:study_environments} shows all environments in our data collection.
\subsubsection{Scenarios}
In each environment, the experimenter inquired participants about different activities they perform on a daily basis.
Participants were presented with two scenarios per environment based on their personal routines for which they should create XR layouts.
Those scenarios covered typical interactions, such as relaxing and workout in the living room; focused work and discussion in the office; and making coffee and cooking with friends in the kitchen.
They created one XR layout for each scenario.

\subsubsection{\system{} Data Collection Tasks}
For interacting with \system, participants completed two different tasks, specifically (1) creating screenshots and (2) generating XR layouts.
\paragraph{\system{} Task 1: Generating app data through screenshots}
Participants collected screenshots of applications using \system{}.
The experimenter instructed them to \textit{``take screenshots of applications that you would find useful for XR within the current environment.''}
Previously taken screenshots (\eg in other environments) were stored and could be reused.
Participants could always update the screenshots if they missed an app or functionality.
They were able to redact personal information, or utilize any of the placeholder applications.

\paragraph{\system{} Task 2: Generating XR widgets and layouts}
For each scenario, participants were asked to create XR widgets with the screenshots they had taken with \system{}.
Participants were instructed that they could \textit{``crop the parts of the app screenshots that they would find useful in XR for the given environment and scenario.''}
They placed the widgets to create an XR layout tailored for the scenario.

\subsection{Procedure}
After completing the consent form and demographic questionnaire, participants were guided to the first environment.
The selection and order of the environments was counterbalanced.
Participants then received a brief training for iOS study app and the Microsoft HoloLens 2, and the overall system.
After the training, participants were presented with the first scenario, collected screenshots, and created the XR layout.
Participants were encouraged to think aloud during the process.
After participants indicated completion, the experimenter introduced the second scenario, and, after completion, guided the participant to the second environment to repeat the data collection for two additional scenarios.
Data collection took between 90 and 120 minutes per participant.
Few participants only had completed three scenarios after 120 minutes, mostly because they created complex layouts, and engaged in extensive think-aloud procedures.
For those participants, we stopped data collection and only completed three scenarios.
We audio-recorded the full sessions for qualitative analysis.
\begin{figure}[t]
    \centering
    \includegraphics[width=\halfwidthfigure]{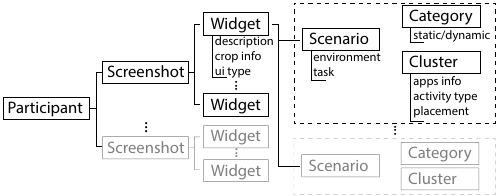}
    \caption{Reference of data and connections that can be found in the \system{} dataset. As indicated by the gray text, participants take multiple screenshots. Participants create multiple widgets by cropping screenshots and placing them. 
    Widgets can be used in multiple scenarios and clusters.}
    \label{fig:ds_classdiagram}
\end{figure}
\subsubsection{Participants and Apparatus}
We recruited 31 paid participants (14 male, 16 female, 1 undisclosed) from a local University through convenience sampling (23) and external participants through a recruiting platform (8).
Participants were aged between 18 and 35 years ($M = 25.9$, $SD = 4.04$), and were students (undergrad and graduate), researchers, or business professionals.
Participants had an average experience with Augmented Reality of $M = 2.16$ ($SD = 1.10$) and Virtual Reality of $M = 2.7$ ($SD = 0.95$), on a scale from 1 (None) to 5 (Expert).
All participants had normal or corrected-to-normal vision based on self-reports.

The first ten participants used the custom \system{} iOS study app on their personal devices (minimum requirement iPhone 10 and iOS 16.0), installed by the experimenter.
The remaining 21 participants took the screenshots on their personal phones, and then transferred them to a dedicated study phone (iPhone 14 Pro) using AirDrop.
This transfer process turned out to be simpler than installing the custom iOS app on their phone, and could be easily repeated on-demand during the study if participants desired other screenshots.
A Microsoft HoloLens 2 was used for delivering the see-through XR experience.
The application platform was used as described in Section~\ref{sec:system}.
\begin{figure*}
    \centering
    \includegraphics[width=0.88\fullwidthfigure]{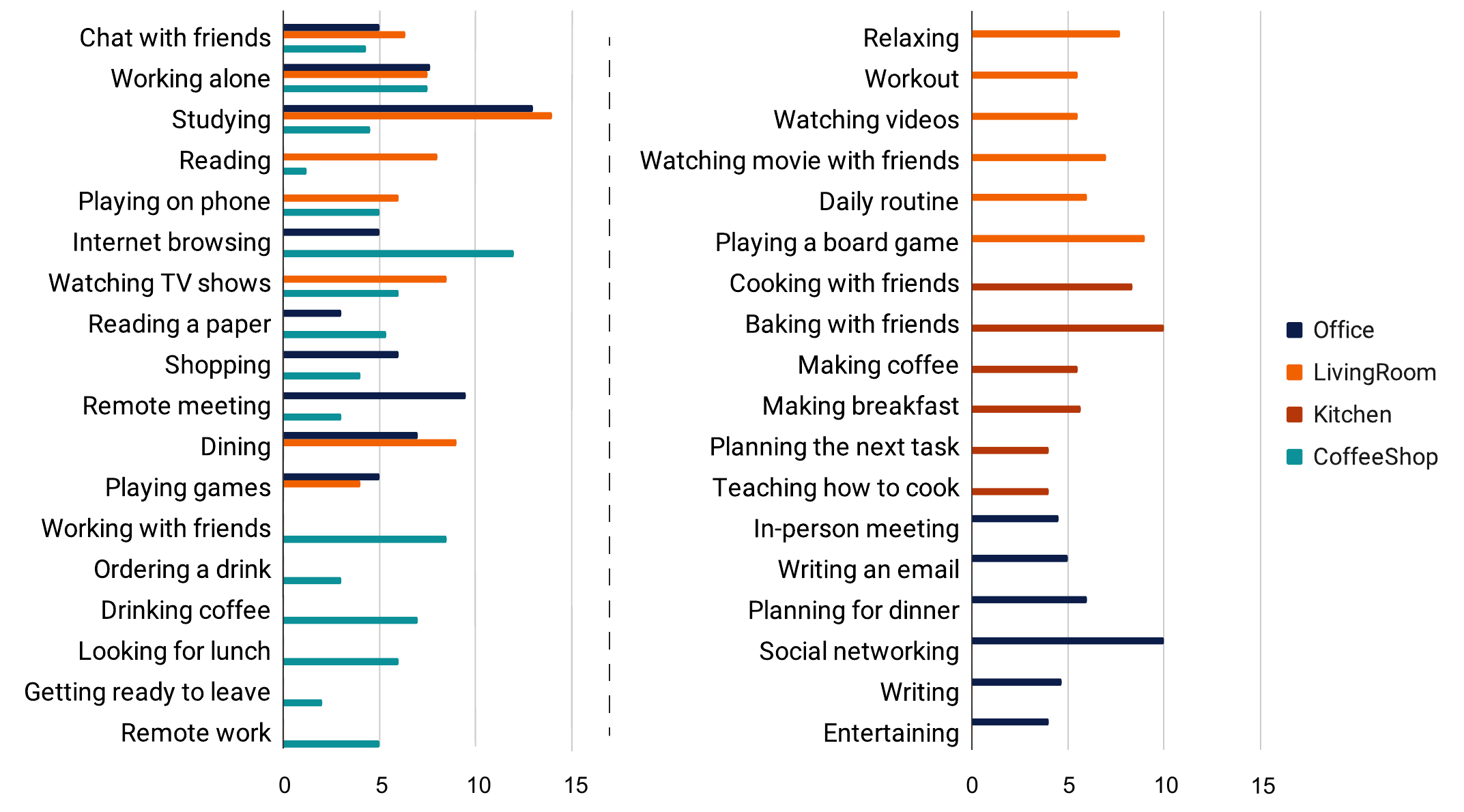}
    \vspace{-1em}
    \caption{Average number of widgets placed in each scenario (environment $\times$ task).}
    \label{fig:widget_distribution_by_scenario}
\end{figure*}
\begin{figure*}
    \centering
    \includegraphics[width=0.65\fullwidthfigure]{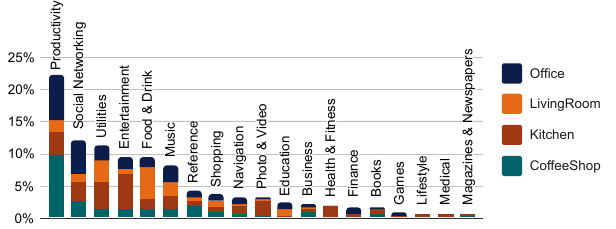}
    \vspace{-0.5em}
    \caption{Distribution of the widgets categories, and across the four different environments.}
    \label{fig:ds_widget_environment}
\end{figure*}
\section{\system{} Dataset}
In the following, we describe the dataset and our annotations. 
The annotations were distributed among four annotators, and cleaned up after completion (\eg unify app names).
The dataset consists of images, raw annotation data stored in the database, and 3D scans of the environments.
Additional analysis was performed in spreadsheets, as all annotations are also available as CSV files.
\subsection{Base Data}
A summary of the annotated data is shown in Figure~\ref{fig:ds_classdiagram}.
Participants took a total of 502 screenshots, on average 16.19 ($SD = 5.90$) per participant, from a total of 178 unique applications and websites.
The \system{} dataset contains 109 unique XR layouts.
Participants created and placed a total of 695 widgets, on average 
22.4 widgets ($SD = 8.29$) per person, and on average 
6.38 widgets ($SD = 3.0$) per scenario. 
The distribution of widgets across scenarios is shown in Figure~\ref{fig:widget_distribution_by_scenario}. Note that the scenarios were selected based on participants' actual daily tasks in the given environment.

The widgets were from 19 different categories (see Figure~\ref{fig:ds_widget_environment}), including 
Productivity (22.2\%, \eg email, work messenger, notes, todo apps),
Social networking (12.1\%, \eg social media, messaging apps),
Utilities (11.3\%, \eg clock, Internet browsers, search engine),
Entertainment (9.6\%, \eg TV shows, movies, video streaming, video-based social media),
Food \& Drink (9.4\%, \eg food delivery, mobile food order apps), and
Music (8.3\%, \eg music-playing, music-streaming apps).

\begin{figure*}[h]
    \centering
    \includegraphics[width=0.98\fullwidthfigure]{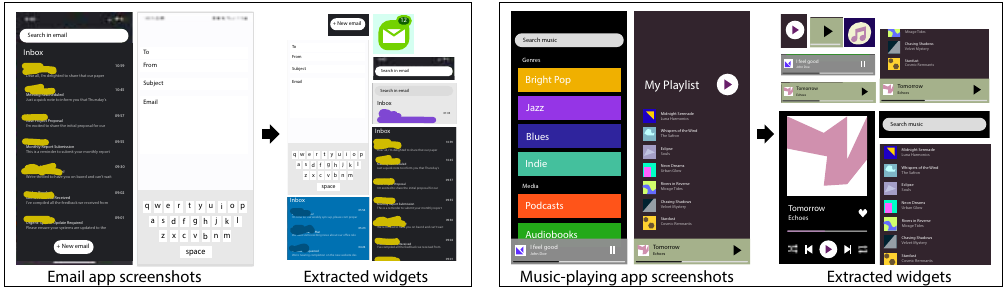}
    \caption{Examples of screenshots and the extracted widgets.}
    \label{fig:cropped_widgets}    
\end{figure*}

Among the 695 widgets, 52.81\% were crops of screenshots into smaller UI element(s) while 47.19\% were used as the whole screenshot.
Examples of an email and music-playing app from various participants are shown in Figure~\ref{fig:cropped_widgets}.
Using the \system{} annotation tools, we annotated the collected widgets.
%
%
The high number of crops highlights that participants appreciate using \textit{individual specialized functionalities}, rather than full applications, and that this, in turn, changes their usage pattern.
This is in contrast to the traditional application-based operating system in desktops and smartphones, which has been adopted in current XR systems as well, \ie users access a digital functionality by opening an application first and navigating inside the app. 
Participants found this navigation step unnecessary in spatial computing and wanted direct access to specific functionalities.

\subsection{Categories of Widgets by Context}
\label{sec:widget_recommendation}
The dataset enables us to provide insights into the usage patterns of XR widgets.
Specifically, we were interested which widgets participants want to use in XR for a given environment, and which widgets they want to use for a given scenario.

\subsubsection{Categories per Environment}
\begin{figure}[h]
    \centering
    \includegraphics[width=\halfwidthfigure]{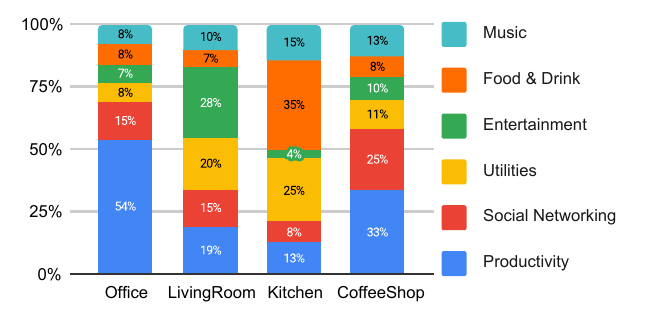}
    \caption{Relative distribution of most frequently used widgets with respect to the individual environments.}
    \label{fig:ds_widget_environment_relative}
\end{figure}

Figure~\ref{fig:ds_widget_environment} shows the different usages for each widget category.
Figure~\ref{fig:ds_widget_environment_relative} provides a more detailed view of the four environments, and how the top-used categories were employed.
All environments yield different utilization in terms of primary category: 
office with 54\% productivity, 
living room with 28\% entertainment, 
kitchen with 35\% food \& drinks (recipes, food order apps as inspiration), 
and coffee shop with 33\% productivity.
All those align well with an intuitive understanding of which apps are useful in particular contexts.

Looking into individual environments also reveals stark differences in general usage patterns, beyond the widgets associated with the primary activity.
In the office environment, the usage of widgets from other categories is relatively equal between 7\% and 15\%, and mostly related to peripheral activities such as social networking and music listening.
In the living room, participants relied more heavily on widgets from utilities (20\%), productivity (19\%), and social networking (15\%).
Utilities were largely ambient interfaces such as clocks and weather apps.
The split between productivity and social networking can be associated with the different activities that participants imagined performing in the environment.
In the kitchen environment, participants similarly leveraged utilities (25\%), as well as music apps (15\%) as ambient displays and peripheral activities (music listening), respectively.
Additionally, participants leveraged productivity apps (13\%) such as notes, email, and calendars.
 Finally, social networking (22\%) and music (13\%) widgets were used frequently for the coffee shop environment.
This shows that environments typically yield one primary widget category based on the current task; but that the composition of the full layout is diverse between contexts. 
The data from the dataset can be used to further automate app suggestions for future XR approaches, for example.
\begin{figure}[h]
    \centering
    \vspace{-0.5em}
    \includegraphics[width=\halfwidthfigure]{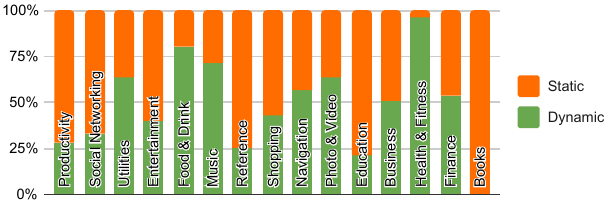}
    \caption{Distribution of most frequently used widgets with respect to whether an environment was static or dynamic.}
    \label{fig:ds_widget_static_dynamic}
\end{figure}
\begin{figure*}[t]
    \centering
    \includegraphics[width=\fullwidthfigure]{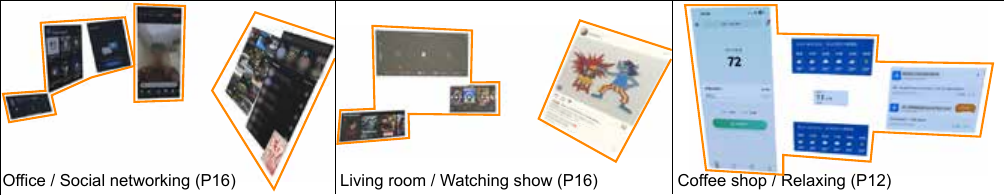}
    \caption{Examples of clusters (\textit{orange border}) for different environments and tasks.}
    \label{fig:clusters-nobackground}
\end{figure*}
\begin{figure}[h]
    \centering
    \includegraphics[width=\halfwidthfigure]{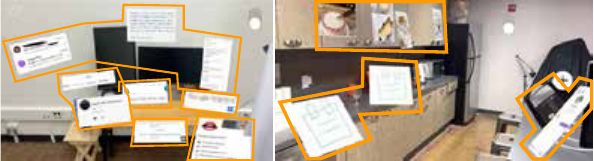}
    \caption{Screenshots of participants' view when creating widgets with annotations of clusters (\textit{orange border}).}
    \label{fig:study_results_clusters}
    \vspace{-1em}
\end{figure}
\begin{figure*}[h]
    \centering
    \includegraphics[width=\fullwidthfigure]{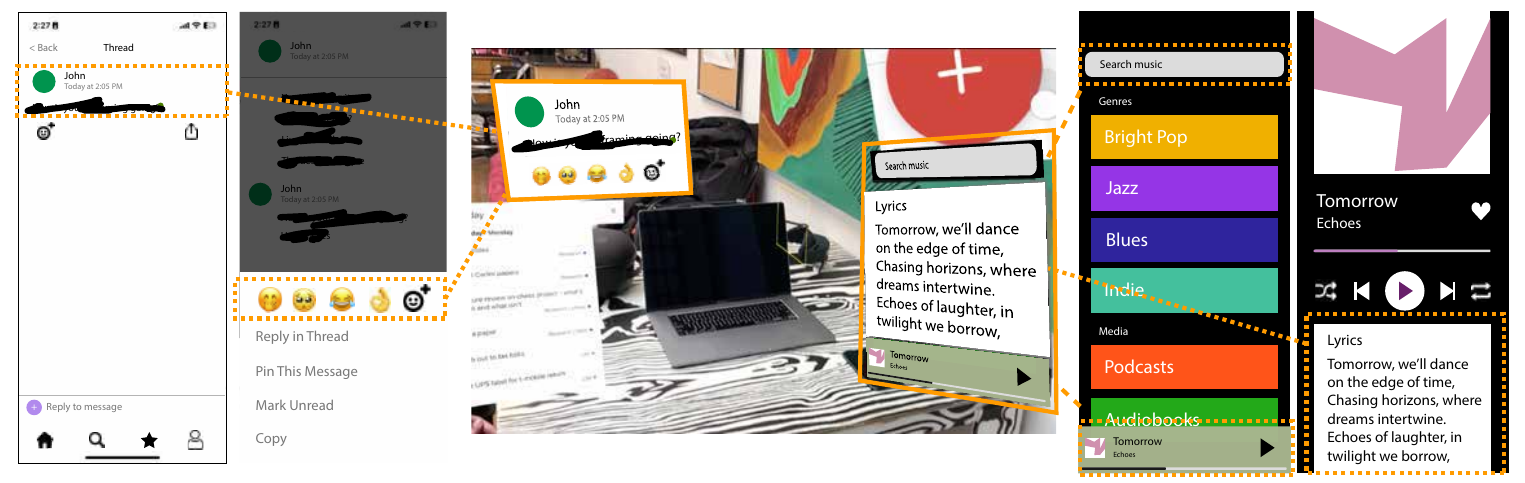}
    \vspace{-2em}
    \caption{Examples of new widget layouts (\textit{solid orange border}) composed from different UI components (\textit{dotted orange border}) of the same app.}
    \label{fig:example_widget_composition}
\end{figure*}
\subsubsection{Categories per Scenarios}
We categorize the activities provided by participants (\eg studying, working) into static (22) and dynamic (30) scenarios. 
Static scenarios are typically single-user scenarios that do not involve a lot of changes in position such as working on a desk.
Dynamic scenarios can involve movement and other objects, \eg cooking, or multiple users.
Figure~\ref{fig:ds_widget_static_dynamic} shows the distribution of the top categories between static and dynamic scenarios.
Books, finance, productivity, entertainment, reference, or education are typically predominantly associated with static environments.
Food \& drinks are largely used in dynamic environments.
Photo \& video, shipping, navigation and business are equally leveraged in both types of scenarios.
\begin{figure*}[t]
    \centering
    \includegraphics[width=0.7\fullwidthfigure]{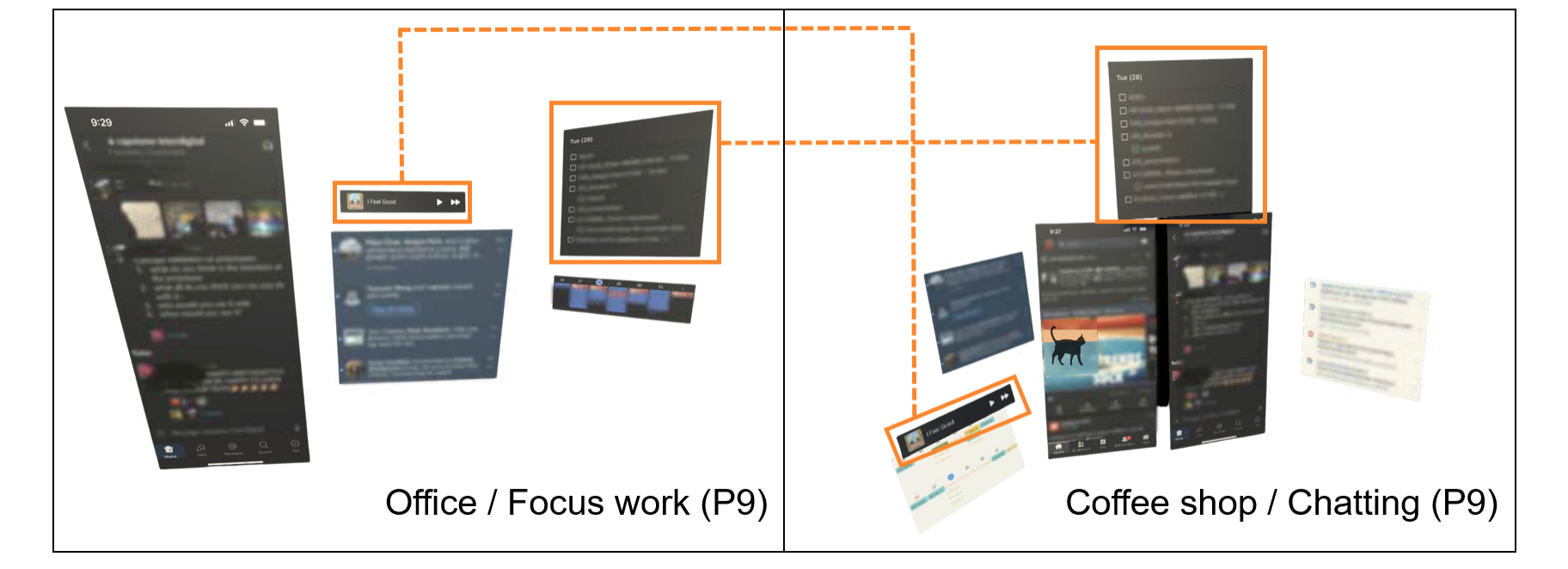}
    \vspace{-0.5em}
    \caption{Examples of the same widgets (\textit{orange border}) reused in clusters for different environments and tasks.}
    \label{fig:widget-recombination}
\end{figure*}
\subsection{Functionality-based Widgets}
We further broke down the data into functionalities to understand \emph{what} functionalities of the categories participants want to use in each environment. 
We annotated descriptions of all 695 widgets using \system{}'s data annotator and grouped widgets by functionalities based on the annotated descriptions. 
We grouped similar functionalities within- and between-applications, \eg `email inbox' in different email apps.
Appendix~\ref{sec:top_functionalities} shows a snippet of these widget annotations---top 10 functionalities from each environment and counts of their occurrences in the environment. 
The app icon and UI navigation are popular functionalities acting as a shortcut to another functionality across all environments.
For other ranked functionalities, different environments exhibit a notably different list of functionalities, which reflects the popular category of each environment.
Our annotated descriptions of the widgets and their functionalities offer a rich understanding of \emph{what} users want to use in XR in different contexts.
We envision that these data will serve as a valuable building block for the future of computational XR, \eg building context-aware XR widget recommendation models.

\subsection{Clusters of XR Widgets} \label{sec:widget_cluster}
Data collected through \system{} also reveals naturally occurring clusters of XR widgets.
We manually annotated the widgets to identify clusters based on spatial distribution.
An annotator viewed each layout, and then clustered the widgets based on proximity and semantics.
Participants created a total of 214 clusters, with an average of 6.90 clusters ($SD = 3.7$) per participant, and an average of 1.96 clusters ($SD = 1.1$) per scenario.
Figure~\ref{fig:clusters-nobackground} shows examples of layouts and annotated clusters based on scenarios.
Figure~\ref{fig:study_results_clusters} shows examples of clusters in different environments.
On average, clusters consisted of 2.98 widgets ($SD = 2.14$).
Participants created new groups of widgets by seamlessly placing individual widgets from different applications together, as exemplified in Figure~\ref{fig:example_widget_composition}.
They also reused the same widgets when forming clusters in various environments for different tasks.
Figure~\ref{fig:widget-recombination} is an example that demonstrates the needs to re-combine functionalities from different applications considering the specific contexts in XR interactions. 

Roughly 30\% of widgets were placed individually, \ie outside of any cluster.
23\% of widgets were placed in pairs.
Of those, roughly half were newly composed layouts from a single application, such as a recipe app that was re-composed of two screens (grocery list, textual description), or a decomposition of the video conferencing application (controls, caller view).
Finally, the remaining 47\% consisted of three or more widgets, with 14\% containing more than 5 widgets.
\subsubsection{Activity Types}
We annotated each cluster to determine whether it is part of the primary activity (\eg email for work; recipe for cooking), peripheral activity (\eg social network for work), or ambient (\eg photos, music, clock).
39.72\% were related to a primary activity, 34.58\% related to a peripheral activity, and 25.7\% related to an ambient display.
Our data shows that participants typically compose their interfaces with one primary application in mind, and enrich their workspace with distributed peripheral information.

\begin{figure}[h]
    \centering
    \includegraphics[width=\halfwidthfigure]{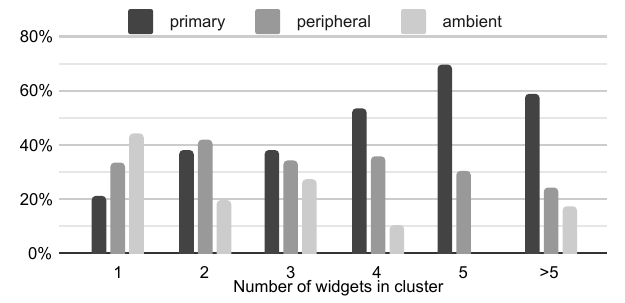}
    \vspace{-0.5em}
    \caption{Distribution of widgets clusters for each activity type, split by the size of each cluster.}
    \label{fig:dataset_cluster_activity}
\end{figure}

The distribution by cluster size is shown in Figure~\ref{fig:dataset_cluster_activity}.
Individual elements were oftentimes used as ambient interfaces, whereas larger clusters were oftentimes composed to resemble primary workspaces (\eg work messenger, email, social media for task \textit{messaging}).
The distribution of peripheral interfaces is roughly constant across the different cluster sizes, at around 30\%.
This outlines a general opportunity for the design of XR interfaces: while the primary activity determines what widgets are used, peripheral interfaces are used across contexts, thus should be available to users at all times.

\subsection{UI Element Types in XR Widgets}
\begin{figure}[h]
    \centering
    \includegraphics[width=\halfwidthfigure]{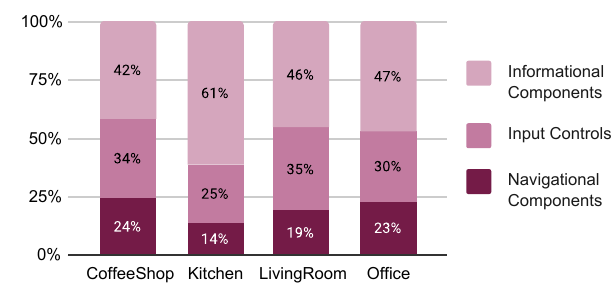}
    \vspace{-0.5em}
    \caption{UI element types of widgets per environment.}
    \label{fig:widget_ui_types}
\end{figure}

We annotated the type(s) of UI elements contained in all widgets based on the UI element types\footnote{https://www.usability.gov/how-to-and-tools/methods/user-interface-elements.html}, namely \textit{Input Controls}, \textit{Navigational Contents}, and \textit{Informational Components}.
This manual annotation process could be further improved by leveraging approaches on smartphone UI understanding~\cite{Wu_Zhang_Nichols_Bigham_2021, Deka_RicoMobile_2017}, depending on the requirements of researchers.
46.74\% contained \textit{Informational Components} that display static or dynamic information, \eg unread emails, currently playing music, food order status, or a virtual plant. \textit{Input Controls} were present in 32.38\% widgets , \eg search box, emoji reactions to a message, or 3D chess game. Lastly,  20.88\% of widgets were \textit{Navigational Components},  and mostly shortcuts to informational components, \eg a work messenger's workspace icon, a button to browse food menu, or a list of shortcuts to project-related documents.
Figure~\ref{fig:widget_ui_types} shows the distribution of different types of UI elements in all widgets for the different environments.
The figure reveals slight differences in usage patterns (\eg more informational components for the kitchen), but are largely comparable across environments.

\subsection{Widget Placement}
In our data collection, we found that participants have different anchoring strategies for different widgets. 
Examples are shown in Figure~\ref{fig:study_results_clusters}.
For example, in work settings, many \textit{Productivity} and \textit{Reference} category widgets were placed with respect to the main workstation. 
P10 wished to save and ``carry'' their workspace layout around the laptop, regardless of the environment, \ie office or coffee shop.
This goes in line with findings from SemanticAdapt~\cite{Cheng_SemanticAdapt_2021} and Lu et al.~\cite{Lu_Xu_2022}.
Some widgets were anchored with respect to fixtures in the room, for example, by the entrance of kitchen, by the window, and on the fridge space.
Existing adaptive XR interfaces focus on changing the placement of XR widgets or applications (\cf Section~\ref{sec:related_work_adaptive_UI}). 
Adaptation algorithms are typically developed based on preliminary data collected in VR environments with predefined virtual interface elements~\cite{Cheng_SemanticAdapt_2021, Lindlbauer_ContextAwareMR_2019}. 
Our approach supplies richer ground truth data for these approaches with a larger set of user-defined widgets.

\section{Discussion}
\system{} enables research to collect rich data on the usage of XR layouts and to draw conclusions about users' preferences, as well as challenges, and opportunities for future XR approaches.
Our dataset is a first step in this direction. 
Besides gathering initial insights, we envision future usage for recommender algorithms and adaptive systems. 
\system{} enables more detailed and personalized findings, that are essential for designing general and adaptive XR interfaces.

\subsection{Design Guidelines}
In the following, we distill findings regarding user preferences for XR systems from our dataset into design guidelines with relevant anecdotal data.
We hope that researchers leverage our dataset and data collection toolset to extend those findings in the future.

\begin{itemize}[leftmargin=*]
    \item \textbf{Design modular UI elements beyond apps for functionality-based XR.} 
    Participants created their layouts by cropping and re-composing existing apps for focused digital access. For example, P5 commented that extracting specific functionalities avoids possible distractions that happen when navigating through their phone. Future XR interfaces should provide widget-based UI elements to facilitate recomposition and clustering by functionality, in addition to following an application-based paradigm. 
    \item \textbf{Develop XR interfaces that adapt to users' space and activity}
    We confirm the hypothesis of context-aware adaptive XR systems~\cite{Lindlbauer_ContextAwareMR_2019, Cheng_SemanticAdapt_2021} that the usage of XR interfaces heavily depends on users' current space and activity, and whether an activity is static or dynamic. 
    P7, for example, chose ``to selectively show and hide information in everyday AR in a way that doesn't feel overwhelming and that complements an existing physical space aesthetically''.
    XR systems should provide context-sensitive information and adapt their usage accordingly.
    \item \textbf{Provide continuous access to peripheral functionalities.}
    More than a third of clusters of XR widgets were used for peripheral tasks such as social media and communication; and were used throughout all environments and scenarios. XR layouts should provide users continuous access to functionalities that might not be relevant to users' primary tasks.
    Especially in work contexts, P2 liked to use the XR space as an ``infinite extended monitor'' (Figure~\ref{fig:study_results_clusters} left) where they could add widgets without having to switch between tabs.
    \item \textbf{Provide seamless integration of ambient information.}
    Users frequently leverage widgets such as clocks or weather as ambient information displays, which should be reflected in any XR layout. This helped P7, for example, to ``not have to switch between apps'' and ``get information from them at a glance without breaking my main task focus'', which supports the Glanceable AR paradigm explored in prior art~\cite{Lu_Bowman_2021,Lu_Davari_Lisle_Li_Bowman_2020,Davari_Lu_Bowman_2022}. Our dataset informs the design for ambient, glanceable interface with grounded data on what widgets were utilized for the ambient use and how they were composed in a layout.
    Ambient XR interface can be further integrated into existing spaces, like decorative physical objects as proposed by Han et al.~\cite{han2023blendmr}.
    \item \textbf{Accommodate both informational components and user input for navigation.} 
    Users leveraged XR for both informational components and input (\eg search, scrolling through emails). XR layouts should provide both functionalities, and not just focus on serving as a flexible display surface. Furthermore, providing users with shortcuts to navigate between functionalities is desirable. 
    \item \textbf{Cluster widgets by functionality for efficient layout composition.} 
    Users frequently created clusters of widgets since they aided in completing a specific task (\eg helpers for cooking), or because they were semantically similar (\eg social media). This insight can be leveraged when assisting users in composing their layouts, for example by providing suggestions, or for algorithms that assist in the placement of virtual elements.
    \item \textbf{Enable personalization mechanisms for various aspects of XR layouts.}    
    While our dataset reveals multiple generalizable patterns, the final designs of participants, unsurprisingly, were highly personal. As with most novel technology, there is no one-size-fits-all interface. Given the flexibility of XR interfaces, and the large diversity in context-dependent factors, efficient mechanisms that enable users to personalize their interfaces are key for building efficient experiences. This personalization should cover multiple aspects of an XR layout, including the selection of functionalities, their placement, and grouping.
\end{itemize}
\vspace{-0.5em}
\subsection{Applications vs. Functionalities for XR}
\system{} enabled participants of the data collection to not only place full applications but to extract widgets that contain specific functionalities through cropping.
This gave them significantly more freedom to create XR layouts that are not bound to current applications.
Participants used this cropping functionality for more than 50\% of the placed widgets, from just cropping unimportant parts, to extracting search bars, email lists, to app icons as shortcuts; or hedonistically to place photos and videos that serve as decorative elements.
We believe that future XR approaches should enable a more integrated experience, and move away from app-based layouts towards more fine-grained functionality-based approaches to provide focused digital access with less distraction, comparable to widgets in modern smartphone operating systems or task-oriented intents in voice user interfaces. 
Prior work on computational understanding of semantic components in mobile UIs~\cite{Wu_Zhang_Nichols_Bigham_2021,Wang_Screen2WordsAutomatic_2021,cho2021reflect} can be used in future works to automate the extraction of functionalities from existing UIs and creation of XR widgets. Some works showcase how these extracted functionalities can be used to generate new interfaces for conversational bots~\cite{Li_Riva_2018}, virtual assistants~\cite{Arsan_Zaidi_Sagar_Kumar_2021}, and customized mashups~\cite{kim2019x,Lee_Kim_Kim_Lee_Kim_Song_Ko_Oh_Shin_2022}.
Data and insights obtained with \system{} serve as groundings for functionality-oriented XR interface design.
\vspace{-0.5em}
\subsection{Different Modes of Data Collection}
We leverage users' personal smartphones in combination with a head-mounted display.
We believe this combination enables users to utilize their personal apps and information, while providing the context of future always-on XR.
Through the headset, users are able to see virtual widgets in space with natural head rotations, without having to point their smartphone at them, which we believe provides higher external validity.
A natural extension to expand the scope of data collection involves enabling users to create widgets not only from smartphones but also from a wider array of personal devices, such as laptops, smartwatches, and tablets.

We decided against a smartphone-only approach, as we believe that the combination with an HMD allows users to fully leverage their environment for creating XR layouts, and limits challenges such as arm fatigue.
Additionally, we decided against performing the data collection in fully immersive VR, since it is unclear whether user behavior is equivalent in the real world.
We believe \system{} leads to more realistic data, since users are in diverse real settings.
We hope to compare VR data collection with our approach in the future, and investigate questions of perceptual and behavioral differences.

\subsection{Privacy-preserving Data Collection}
We provide users with two privacy-preserving mechanisms to share data: they can blur personal information, and we provide placeholder apps (\eg email, dating).
While those mitigate some privacy concerns, the nature of the data collection (\ie with an experimenter) poses challenges.
Users might, for example, not be comfortable sharing specific apps they would like to use in an environment.
Our dataset required manual cleanup to blur text as an additional step.
A combination with crowd-sourcing might provide a balance in terms of privacy and data.
Our platform collects the 3D position of the placed widgets and does not collect photos or videos.
We extend our data with static 3D scans or abstracted 3D semantic maps.
This approach enables researchers to gather insights into the placement and grouping of specific functionalities without sacrificing privacy.

\subsection{Limitations \& Future Work}
\subsubsection{Dynamic UI Layouts}
Our data collection tool currently supports only static, world-centric placement. 
The placed widgets are affixed to the world. The tool supports updating the widget position but does not support a dynamic layout.
In our current data collection, participants created static UI layouts to complete their tasks, and they found these layouts to be sufficient, for example, to simulate a user-centric widget by placing the widget around themselves.
However, the current dataset does not annotate whether the widgets were anchored to the world or the user.
We regard it as interesting future work to explore the design of temporally dynamic UI layouts, where users can specify user-centric widget behaviors.
These dynamic layouts can be especially helpful for tasks involving multiple steps that need to be completed sequentially.
For example, when cooking in the kitchen, the user may need different functionalities at different steps (choosing from recipes, following tutorials).
They should be able to adjust the layout dynamically to suit their needs.
With actionable adaptation of \system{}, for example letting users define both the location and the temporal order of widgets when placing them, we believe we can facilitate the exploration on dynamic UI layouts in the future.

\subsubsection{Interactivity}
Our current platform leverages screenshots rather than functioning virtual elements.
Therefore, it remains unknown how users would actually interact with these interfaces, and how the users would modify and adapt the interfaces in turn through the course of interactions.
Ideally, a personalized layout would provide users with fully interactive versions to study long-term use and deployment.
Although our dataset and initial analyses offer a warm start for the future of adaptive XR interfaces, it is still an open question how users would interact with these UIs and how the interactions would affect the UIs in turn.

\subsubsection{In-situ data collection using \system}
\system{}'s in-situ approach enables collecting data for in-depth and comprehensive understandings of XR experiences, as showcased in our dataset.
Our data collection took approximately two hours for participants to take screenshots and create four XR layouts in two environments under a think-aloud protocol, including introduction and debriefing.
This constitutes considerable effort, especially compared to online studies~\cite{dudley2021crowdsourcing, Lee2022}.
We believe the development of an in-the-wild, interactive version of \system{} that enables more longitudinal data collection provides an interesting direction for future work.

\subsubsection{\system{} for Computational XR}
We see a core usage of our dataset and \system{} in adaptive XR systems.
We envision a wide range of usages, from providing interface recommendations to users based on previously selected functionalities, to automatically extracting widgets and functionalities, to automatic layout approaches.
Our data is the first that comes close to providing ground-truth data for XR systems.
For example, our data can be used to validate previous findings on XR layout placement strategies, \eg placing near objects with semantic relevance~\cite{Cheng_SemanticAdapt_2021}, and reveal emerging placement strategies from a more in-depth analysis in relation to the environment. These newly discovered strategies can be in turn incorporated as additional factors or constraints in optimization objectives used in adaptive XR interface works~(\eg\cite{Lindlbauer_ContextAwareMR_2019, Cheng_SemanticAdapt_2021, fender2018optispace}).
Furthermore, findings from previous works can be incorporated in MineXR data collection process as well. For example, widget extraction can be facilitated with automated UI structure extraction~\cite{Wu_Zhang_Nichols_Bigham_2021}. Widget placement can be facilitated by snapping to planes~\cite{Nuernberger_Ofek_Benko_Wilson_2016}, snapping to semantically-related objects~\cite{Cheng_SemanticAdapt_2021}, or integrating ambient information with physical objects~\cite{han2023blendmr}.
We hope that by open-sourcing our dataset as well as our data collection and analysis platform, researchers will further expand and utilize our dataset.

\section{Conclusion}
We present \system, a novel data collection platform that enables researchers to collect data on users' preferences in terms of layout and desired functionality of future XR interfaces and XR experiences.
We leverage participants' personal smartphones for the selection and placement of widgets, and a head-mounted display for presenting XR content.
Our platform enables in-situ data collection and provides researchers with the opportunity to create rich datasets.
We contribute a dataset of 109 unique XR layouts from 31 users composed of 695 widgets, as well as an analysis that led to a set of design guidelines for XR interfaces. 
We believe that \system{} is a main step towards designing future XR interfaces that are personalized and beneficial for users. 
Our insights benefit manual design approaches as well as adaptive XR systems.
\system{} tools and dataset are available at \textit{\url{https://augmented-perception.org/publications/2024-minexr.html}}.

\begin{acks}
We thank Kaitlyn Ng and Yixuan (Sherry) Wang for their assistance with running user studies and annotating data.
Funding to attend this conference was provided by the CMU GSA/Provost Conference Funding.
\end{acks}

\bibliographystyle{ACM-Reference-Format}
\bibliography{references}


\begin{thebibliography}{81}


\ifx \showCODEN    \undefined \def \showCODEN     #1{\unskip}     \fi
\ifx \showDOI      \undefined \def \showDOI       #1{#1}\fi
\ifx \showISBNx    \undefined \def \showISBNx     #1{\unskip}     \fi
\ifx \showISBNxiii \undefined \def \showISBNxiii  #1{\unskip}     \fi
\ifx \showISSN     \undefined \def \showISSN      #1{\unskip}     \fi
\ifx \showLCCN     \undefined \def \showLCCN      #1{\unskip}     \fi
\ifx \shownote     \undefined \def \shownote      #1{#1}          \fi
\ifx \showarticletitle \undefined \def \showarticletitle #1{#1}   \fi
\ifx \showURL      \undefined \def \showURL       {\relax}        \fi
\providecommand\bibfield[2]{#2}
\providecommand\bibinfo[2]{#2}
\providecommand\natexlab[1]{#1}
\providecommand\showeprint[2][]{arXiv:#2}

\bibitem[Arsan et~al\mbox{.}(2021)]%
        {Arsan_Zaidi_Sagar_Kumar_2021}
\bibfield{author}{\bibinfo{person}{Deniz Arsan}, \bibinfo{person}{Ali Zaidi},
  \bibinfo{person}{Aravind Sagar}, {and} \bibinfo{person}{Ranjitha Kumar}.}
  \bibinfo{year}{2021}\natexlab{}.
\newblock \showarticletitle{App-Based Task Shortcuts for Virtual Assistants}.
  In \bibinfo{booktitle}{\emph{The 34th Annual ACM Symposium on User Interface
  Software and Technology}}. \bibinfo{publisher}{ACM},
  \bibinfo{address}{Virtual Event USA}, \bibinfo{pages}{1089–1099}.
\newblock
\showISBNx{978-1-4503-8635-7}
\urldef\tempurl%
\url{https://doi.org/10.1145/3472749.3474808}
\showDOI{\tempurl}


\bibitem[Azuma and Furmanski(2003)]%
        {azuma2003evaluating}
\bibfield{author}{\bibinfo{person}{Ronald Azuma} {and} \bibinfo{person}{Chris
  Furmanski}.} \bibinfo{year}{2003}\natexlab{}.
\newblock \showarticletitle{Evaluating label placement for augmented reality
  view management}. In \bibinfo{booktitle}{\emph{The Second IEEE and ACM
  International Symposium on Mixed and Augmented Reality, 2003. Proceedings.}}
  IEEE, \bibinfo{pages}{66--75}.
\newblock


\bibitem[Bai et~al\mbox{.}(2021)]%
        {Bai_Zang_Xu_Sunkara_Rastogi_Chen_Arcas_2021}
\bibfield{author}{\bibinfo{person}{Chongyang Bai}, \bibinfo{person}{Xiaoxue
  Zang}, \bibinfo{person}{Ying Xu}, \bibinfo{person}{Srinivas Sunkara},
  \bibinfo{person}{Abhinav Rastogi}, \bibinfo{person}{Jindong Chen}, {and}
  \bibinfo{person}{Blaise Aguera~y Arcas}.} \bibinfo{year}{2021}\natexlab{}.
\newblock \showarticletitle{UIBert: Learning Generic Multimodal Representations
  for UI Understanding}.
\newblock  \bibinfo{number}{arXiv:2107.13731} (\bibinfo{date}{Aug}
  \bibinfo{year}{2021}).
\newblock
\urldef\tempurl%
\url{http://arxiv.org/abs/2107.13731}
\showURL{%
\tempurl}
\newblock
\shownote{arXiv:2107.13731 [cs]}.


\bibitem[Bell et~al\mbox{.}(2001)]%
        {bell2001view}
\bibfield{author}{\bibinfo{person}{Blaine Bell}, \bibinfo{person}{Steven
  Feiner}, {and} \bibinfo{person}{Tobias H{\"o}llerer}.}
  \bibinfo{year}{2001}\natexlab{}.
\newblock \showarticletitle{View management for virtual and augmented reality}.
  In \bibinfo{booktitle}{\emph{Proceedings of the 14th annual ACM symposium on
  User interface software and technology}}. \bibinfo{pages}{101--110}.
\newblock


\bibitem[Belo and Lystb(2022)]%
        {Belo_Lystb_2022}
\bibfield{author}{\bibinfo{person}{João Belo} {and} \bibinfo{person}{Mathias~N
  Lystb}.} \bibinfo{year}{2022}\natexlab{}.
\newblock \showarticletitle{AUIT – the Adaptive User Interfaces Toolkit for
  Designing XR Applications}.
\newblock  (\bibinfo{year}{2022}), \bibinfo{pages}{17}.
\newblock


\bibitem[Brudy et~al\mbox{.}(2019)]%
        {Brudy2019cross_device}
\bibfield{author}{\bibinfo{person}{Frederik Brudy}, \bibinfo{person}{Christian
  Holz}, \bibinfo{person}{Roman R{\"a}dle}, \bibinfo{person}{Chi-Jui Wu},
  \bibinfo{person}{Steven Houben}, \bibinfo{person}{Clemens~Nylandsted
  Klokmose}, {and} \bibinfo{person}{Nicolai Marquardt}.}
  \bibinfo{year}{2019}\natexlab{}.
\newblock \showarticletitle{Cross-device taxonomy: Survey, opportunities and
  challenges of interactions spanning across multiple devices}. In
  \bibinfo{booktitle}{\emph{Proceedings of the 2019 {CHI} Conference on Human
  Factors in Computing Systems}} (Glasgow Scotland Uk).
  \bibinfo{publisher}{ACM}, \bibinfo{address}{New York, NY, USA}.
\newblock
\showISBNx{9781450359702}
\urldef\tempurl%
\url{https://doi.org/10.1145/3290605.3300792}
\showDOI{\tempurl}


\bibitem[Cheng et~al\mbox{.}(2021)]%
        {Cheng_SemanticAdapt_2021}
\bibfield{author}{\bibinfo{person}{Yifei Cheng}, \bibinfo{person}{Yukang Yan},
  \bibinfo{person}{Xin Yi}, \bibinfo{person}{Yuanchun Shi}, {and}
  \bibinfo{person}{David Lindlbauer}.} \bibinfo{year}{2021}\natexlab{}.
\newblock \showarticletitle{SemanticAdapt: Optimization-based Adaptation of
  Mixed Reality Layouts Leveraging Virtual-Physical Semantic Connections}. In
  \bibinfo{booktitle}{\emph{The 34th Annual ACM Symposium on User Interface
  Software and Technology}}. \bibinfo{publisher}{ACM},
  \bibinfo{address}{Virtual Event USA}, \bibinfo{pages}{282–297}.
\newblock
\showISBNx{978-1-4503-8635-7}
\urldef\tempurl%
\url{https://doi.org/10.1145/3472749.3474750}
\showDOI{\tempurl}


\bibitem[Cho et~al\mbox{.}(2021)]%
        {cho2021reflect}
\bibfield{author}{\bibinfo{person}{Hyunsung Cho}, \bibinfo{person}{DaEun Choi},
  \bibinfo{person}{Donghwi Kim}, \bibinfo{person}{Wan~Ju Kang},
  \bibinfo{person}{Eun~Kyoung Choe}, {and} \bibinfo{person}{Sung-Ju Lee}.}
  \bibinfo{year}{2021}\natexlab{}.
\newblock \showarticletitle{Reflect, not regret: Understanding regretful
  smartphone use with app feature-level analysis}.
\newblock \bibinfo{journal}{\emph{Proceedings of the ACM on Human-Computer
  Interaction}} \bibinfo{volume}{5}, \bibinfo{number}{CSCW2}
  (\bibinfo{year}{2021}), \bibinfo{pages}{1--36}.
\newblock


\bibitem[Davari et~al\mbox{.}(2022)]%
        {Davari_Lu_Bowman_2022}
\bibfield{author}{\bibinfo{person}{Shakiba Davari}, \bibinfo{person}{Feiyu Lu},
  {and} \bibinfo{person}{Doug~A. Bowman}.} \bibinfo{year}{2022}\natexlab{}.
\newblock \showarticletitle{Validating the Benefits of Glanceable and
  Context-Aware Augmented Reality for Everyday Information Access Tasks}. In
  \bibinfo{booktitle}{\emph{2022 IEEE Conference on Virtual Reality and 3D User
  Interfaces (VR)}}. \bibinfo{publisher}{IEEE}, \bibinfo{address}{Christchurch,
  New Zealand}, \bibinfo{pages}{436–444}.
\newblock
\showISBNx{978-1-66549-617-9}
\urldef\tempurl%
\url{https://doi.org/10.1109/VR51125.2022.00063}
\showDOI{\tempurl}


\bibitem[Deka et~al\mbox{.}(2017a)]%
        {Deka_RicoMobile_2017}
\bibfield{author}{\bibinfo{person}{Biplab Deka}, \bibinfo{person}{Zifeng
  Huang}, \bibinfo{person}{Chad Franzen}, \bibinfo{person}{Joshua Hibschman},
  \bibinfo{person}{Daniel Afergan}, \bibinfo{person}{Yang Li},
  \bibinfo{person}{Jeffrey Nichols}, {and} \bibinfo{person}{Ranjitha Kumar}.}
  \bibinfo{year}{2017}\natexlab{a}.
\newblock \showarticletitle{Rico: {{A Mobile App Dataset}} for {{Building
  Data-Driven Design Applications}}}. In \bibinfo{booktitle}{\emph{Proceedings
  of the 30th {{Annual ACM Symposium}} on {{User Interface Software}} and
  {{Technology}}}}. \bibinfo{publisher}{{ACM}}, \bibinfo{address}{{Qu\'ebec
  City QC Canada}}, \bibinfo{pages}{845--854}.
\newblock
\showISBNx{978-1-4503-4981-9}
\urldef\tempurl%
\url{https://doi.org/10.1145/3126594.3126651}
\showDOI{\tempurl}


\bibitem[Deka et~al\mbox{.}(2017b)]%
        {Deka_ZIPTZeroIntegration_2017}
\bibfield{author}{\bibinfo{person}{Biplab Deka}, \bibinfo{person}{Zifeng
  Huang}, \bibinfo{person}{Chad Franzen}, \bibinfo{person}{Jeffrey Nichols},
  \bibinfo{person}{Yang Li}, {and} \bibinfo{person}{Ranjitha Kumar}.}
  \bibinfo{year}{2017}\natexlab{b}.
\newblock \showarticletitle{{{ZIPT}}: {{Zero-Integration Performance Testing}}
  of {{Mobile App Designs}}}. In \bibinfo{booktitle}{\emph{Proceedings of the
  30th {{Annual ACM Symposium}} on {{User Interface Software}} and
  {{Technology}}}}. \bibinfo{publisher}{{ACM}}, \bibinfo{address}{{Qu\'ebec
  City QC Canada}}, \bibinfo{pages}{727--736}.
\newblock
\showISBNx{978-1-4503-4981-9}
\urldef\tempurl%
\url{https://doi.org/10.1145/3126594.3126647}
\showDOI{\tempurl}


\bibitem[Deka et~al\mbox{.}(2016)]%
        {Deka_ERICA_2016}
\bibfield{author}{\bibinfo{person}{Biplab Deka}, \bibinfo{person}{Zifeng
  Huang}, {and} \bibinfo{person}{Ranjitha Kumar}.}
  \bibinfo{year}{2016}\natexlab{}.
\newblock \showarticletitle{ERICA: Interaction Mining Mobile Apps}. In
  \bibinfo{booktitle}{\emph{Proceedings of the 29th Annual Symposium on User
  Interface Software and Technology}}. \bibinfo{publisher}{ACM},
  \bibinfo{address}{Tokyo Japan}, \bibinfo{pages}{767–776}.
\newblock
\showISBNx{978-1-4503-4189-9}
\urldef\tempurl%
\url{https://doi.org/10.1145/2984511.2984581}
\showDOI{\tempurl}


\bibitem[DiVerdi et~al\mbox{.}(2004)]%
        {diverdi2004level}
\bibfield{author}{\bibinfo{person}{Stephen DiVerdi}, \bibinfo{person}{Tobias
  Hollerer}, {and} \bibinfo{person}{Richard Schreyer}.}
  \bibinfo{year}{2004}\natexlab{}.
\newblock \showarticletitle{Level of detail interfaces}. In
  \bibinfo{booktitle}{\emph{Third IEEE and ACM International Symposium on Mixed
  and Augmented Reality}}. IEEE, \bibinfo{pages}{300--301}.
\newblock


\bibitem[Dong et~al\mbox{.}(2021)]%
        {dong2021tailored}
\bibfield{author}{\bibinfo{person}{Zhi-Chao Dong}, \bibinfo{person}{Wenming
  Wu}, \bibinfo{person}{Zenghao Xu}, \bibinfo{person}{Qi Sun},
  \bibinfo{person}{Guanjie Yuan}, \bibinfo{person}{Ligang Liu}, {and}
  \bibinfo{person}{Xiao-Ming Fu}.} \bibinfo{year}{2021}\natexlab{}.
\newblock \showarticletitle{Tailored reality: Perception-aware scene
  restructuring for adaptive vr navigation}.
\newblock \bibinfo{journal}{\emph{ACM Transactions on Graphics (TOG)}}
  \bibinfo{volume}{40}, \bibinfo{number}{5} (\bibinfo{year}{2021}),
  \bibinfo{pages}{1--15}.
\newblock


\bibitem[Du et~al\mbox{.}(2020)]%
        {du2020depthlab}
\bibfield{author}{\bibinfo{person}{Ruofei Du}, \bibinfo{person}{Eric Turner},
  \bibinfo{person}{Maksym Dzitsiuk}, \bibinfo{person}{Luca Prasso},
  \bibinfo{person}{Ivo Duarte}, \bibinfo{person}{Jason Dourgarian},
  \bibinfo{person}{Joao Afonso}, \bibinfo{person}{Jose Pascoal},
  \bibinfo{person}{Josh Gladstone}, \bibinfo{person}{Nuno Cruces},
  {et~al\mbox{.}}} \bibinfo{year}{2020}\natexlab{}.
\newblock \showarticletitle{DepthLab: Real-time 3D interaction with depth maps
  for mobile augmented reality}. In \bibinfo{booktitle}{\emph{Proceedings of
  the 33rd Annual ACM Symposium on User Interface Software and Technology}}.
  \bibinfo{pages}{829--843}.
\newblock


\bibitem[Dudley et~al\mbox{.}(2021)]%
        {dudley2021crowdsourcing}
\bibfield{author}{\bibinfo{person}{John~J Dudley}, \bibinfo{person}{Jason~T
  Jacques}, {and} \bibinfo{person}{Per~Ola Kristensson}.}
  \bibinfo{year}{2021}\natexlab{}.
\newblock \showarticletitle{Crowdsourcing design guidance for contextual
  adaptation of text content in augmented reality}. In
  \bibinfo{booktitle}{\emph{Proceedings of the 2021 CHI Conference on Human
  Factors in Computing Systems}}. \bibinfo{pages}{1--14}.
\newblock


\bibitem[Ens et~al\mbox{.}(2017)]%
        {Ens_Ivy_2017}
\bibfield{author}{\bibinfo{person}{Barrett Ens}, \bibinfo{person}{Fraser
  Anderson}, \bibinfo{person}{Tovi Grossman}, \bibinfo{person}{Michelle
  Annett}, \bibinfo{person}{Pourang Irani}, {and} \bibinfo{person}{George
  Fitzmaurice}.} \bibinfo{year}{2017}\natexlab{}.
\newblock \showarticletitle{Ivy: Exploring Spatially Situated Visual
  Programming for Authoring and Understanding Intelligent Environments}. In
  \bibinfo{booktitle}{\emph{Proceedings of the 43rd Graphics Interface
  Conference}} (Edmonton, Alberta, Canada) \emph{(\bibinfo{series}{GI '17})}.
  \bibinfo{publisher}{Canadian Human-Computer Communications Society},
  \bibinfo{address}{Waterloo, CAN}, \bibinfo{pages}{156–162}.
\newblock
\showISBNx{9780994786821}


\bibitem[Ens et~al\mbox{.}(2015)]%
        {Ens_Ofek_Bruce_Irani_2015}
\bibfield{author}{\bibinfo{person}{Barrett Ens}, \bibinfo{person}{Eyal Ofek},
  \bibinfo{person}{Neil Bruce}, {and} \bibinfo{person}{Pourang Irani}.}
  \bibinfo{year}{2015}\natexlab{}.
\newblock \showarticletitle{Spatial Constancy of Surface-Embedded Layouts
  across Multiple Environments}. In \bibinfo{booktitle}{\emph{Proceedings of
  the 3rd ACM Symposium on Spatial User Interaction}}.
  \bibinfo{publisher}{ACM}, \bibinfo{address}{Los Angeles California USA},
  \bibinfo{pages}{65–68}.
\newblock
\showISBNx{978-1-4503-3703-8}
\urldef\tempurl%
\url{https://doi.org/10.1145/2788940.2788954}
\showDOI{\tempurl}


\bibitem[Fender et~al\mbox{.}(2018)]%
        {fender2018optispace}
\bibfield{author}{\bibinfo{person}{Andreas Fender}, \bibinfo{person}{Philipp
  Herholz}, \bibinfo{person}{Marc Alexa}, {and} \bibinfo{person}{J{\"o}rg
  M{\"u}ller}.} \bibinfo{year}{2018}\natexlab{}.
\newblock \showarticletitle{Optispace: Automated placement of interactive 3d
  projection mapping content}. In \bibinfo{booktitle}{\emph{Proceedings of the
  2018 CHI Conference on Human Factors in Computing Systems}}.
  \bibinfo{pages}{1--11}.
\newblock


\bibitem[Fender et~al\mbox{.}(2017)]%
        {fender2017heatspace}
\bibfield{author}{\bibinfo{person}{Andreas Fender}, \bibinfo{person}{David
  Lindlbauer}, \bibinfo{person}{Philipp Herholz}, \bibinfo{person}{Marc Alexa},
  {and} \bibinfo{person}{J{\"o}rg M{\"u}ller}.}
  \bibinfo{year}{2017}\natexlab{}.
\newblock \showarticletitle{Heatspace: Automatic placement of displays by
  empirical analysis of user behavior}. In
  \bibinfo{booktitle}{\emph{Proceedings of the 30th Annual ACM Symposium on
  User Interface Software and Technology}}. \bibinfo{pages}{611--621}.
\newblock


\bibitem[Fischer et~al\mbox{.}(2018)]%
        {Fischer_Campagna_Xu_Lam_2018}
\bibfield{author}{\bibinfo{person}{Michael Fischer}, \bibinfo{person}{Giovanni
  Campagna}, \bibinfo{person}{Silei Xu}, {and} \bibinfo{person}{Monica~S.
  Lam}.} \bibinfo{year}{2018}\natexlab{}.
\newblock \showarticletitle{Brassau: automatic generation of graphical user
  interfaces for virtual assistants}. In \bibinfo{booktitle}{\emph{Proceedings
  of the 20th International Conference on Human-Computer Interaction with
  Mobile Devices and Services}}. \bibinfo{publisher}{ACM},
  \bibinfo{address}{Barcelona Spain}, \bibinfo{pages}{1–12}.
\newblock
\showISBNx{978-1-4503-5898-9}
\urldef\tempurl%
\url{https://doi.org/10.1145/3229434.3229481}
\showDOI{\tempurl}


\bibitem[Gajos and Weld(2004)]%
        {gajos2004supple}
\bibfield{author}{\bibinfo{person}{Krzysztof Gajos} {and}
  \bibinfo{person}{Daniel~S Weld}.} \bibinfo{year}{2004}\natexlab{}.
\newblock \showarticletitle{SUPPLE: automatically generating user interfaces}.
  In \bibinfo{booktitle}{\emph{Proceedings of the 9th international conference
  on Intelligent user interfaces}}. \bibinfo{pages}{93--100}.
\newblock


\bibitem[Gal et~al\mbox{.}(2014)]%
        {gal2014flare}
\bibfield{author}{\bibinfo{person}{Ran Gal}, \bibinfo{person}{Lior Shapira},
  \bibinfo{person}{Eyal Ofek}, {and} \bibinfo{person}{Pushmeet Kohli}.}
  \bibinfo{year}{2014}\natexlab{}.
\newblock \showarticletitle{FLARE: Fast layout for augmented reality
  applications}. In \bibinfo{booktitle}{\emph{2014 IEEE international symposium
  on mixed and augmented reality (ISMAR)}}. IEEE, \bibinfo{pages}{207--212}.
\newblock


\bibitem[Gasques et~al\mbox{.}(2019)]%
        {Gasques_WhatYou_2019}
\bibfield{author}{\bibinfo{person}{Danilo Gasques}, \bibinfo{person}{Janet~G.
  Johnson}, \bibinfo{person}{Tommy Sharkey}, {and} \bibinfo{person}{Nadir
  Weibel}.} \bibinfo{year}{2019}\natexlab{}.
\newblock \showarticletitle{What {{You Sketch Is What You Get}}: {{Quick}} and
  {{Easy Augmented Reality Prototyping}} with {{PintAR}}}. In
  \bibinfo{booktitle}{\emph{Extended {{Abstracts}} of the 2019 {{CHI
  Conference}} on {{Human Factors}} in {{Computing Systems}}}}.
  \bibinfo{publisher}{{ACM}}, \bibinfo{address}{{Glasgow Scotland Uk}},
  \bibinfo{pages}{1--6}.
\newblock
\showISBNx{978-1-4503-5971-9}
\urldef\tempurl%
\url{https://doi.org/10.1145/3290607.3312847}
\showDOI{\tempurl}


\bibitem[Ghouaiel et~al\mbox{.}(2014)]%
        {ghouaiel2014adaptive}
\bibfield{author}{\bibinfo{person}{Nehla Ghouaiel}, \bibinfo{person}{Jean-Marc
  Cieutat}, {and} \bibinfo{person}{Jean-Pierre Jessel}.}
  \bibinfo{year}{2014}\natexlab{}.
\newblock \showarticletitle{Adaptive augmented reality: plasticity of
  augmentations}. In \bibinfo{booktitle}{\emph{Proceedings of the 2014 Virtual
  Reality International Conference}}. \bibinfo{pages}{1--4}.
\newblock


\bibitem[Grasset et~al\mbox{.}(2012)]%
        {grasset2012image}
\bibfield{author}{\bibinfo{person}{Raphael Grasset}, \bibinfo{person}{Tobias
  Langlotz}, \bibinfo{person}{Denis Kalkofen}, \bibinfo{person}{Markus
  Tatzgern}, {and} \bibinfo{person}{Dieter Schmalstieg}.}
  \bibinfo{year}{2012}\natexlab{}.
\newblock \showarticletitle{Image-driven view management for augmented reality
  browsers}. In \bibinfo{booktitle}{\emph{2012 IEEE International Symposium on
  Mixed and Augmented Reality (ISMAR)}}. IEEE, \bibinfo{pages}{177--186}.
\newblock


\bibitem[Guven and Feiner(2003)]%
        {Guven_Authoring3D_2003}
\bibfield{author}{\bibinfo{person}{Sinem Guven} {and} \bibinfo{person}{Steven
  Feiner}.} \bibinfo{year}{2003}\natexlab{}.
\newblock \showarticletitle{Authoring {{3D}} Hypermedia for Wearable Augmented
  and Virtual Reality}. In \bibinfo{booktitle}{\emph{Seventh {{IEEE
  International Symposium}} on {{Wearable Computers}}, 2003. {{Proceedings}}.}}
  \bibinfo{publisher}{{IEEE}}, \bibinfo{address}{{White Plains, NY, USA}},
  \bibinfo{pages}{118--126}.
\newblock
\showISBNx{978-0-7695-2034-6}
\urldef\tempurl%
\url{https://doi.org/10.1109/ISWC.2003.1241401}
\showDOI{\tempurl}


\bibitem[Guven et~al\mbox{.}(2006)]%
        {Guven_MobileAugmented_2006}
\bibfield{author}{\bibinfo{person}{Sinem Guven}, \bibinfo{person}{Steven
  Feiner}, {and} \bibinfo{person}{Ohan Oda}.} \bibinfo{year}{2006}\natexlab{}.
\newblock \showarticletitle{Mobile Augmented Reality Interaction Techniques for
  Authoring Situated Media On-Site}. In \bibinfo{booktitle}{\emph{2006
  {{IEEE}}/{{ACM International Symposium}} on {{Mixed}} and {{Augmented
  Reality}}}}. \bibinfo{publisher}{{IEEE}}, \bibinfo{address}{{Santa Barbara,
  CA, USA}}, \bibinfo{pages}{235--236}.
\newblock
\showISBNx{978-1-4244-0650-0 978-1-4244-0651-7}
\urldef\tempurl%
\url{https://doi.org/10.1109/ISMAR.2006.297821}
\showDOI{\tempurl}


\bibitem[Han et~al\mbox{.}(2020)]%
        {han2020live}
\bibfield{author}{\bibinfo{person}{Lei Han}, \bibinfo{person}{Tian Zheng},
  \bibinfo{person}{Yinheng Zhu}, \bibinfo{person}{Lan Xu}, {and}
  \bibinfo{person}{Lu Fang}.} \bibinfo{year}{2020}\natexlab{}.
\newblock \showarticletitle{Live semantic 3d perception for immersive augmented
  reality}.
\newblock \bibinfo{journal}{\emph{IEEE transactions on visualization and
  computer graphics}} \bibinfo{volume}{26}, \bibinfo{number}{5}
  (\bibinfo{year}{2020}), \bibinfo{pages}{2012--2022}.
\newblock


\bibitem[Han et~al\mbox{.}(2023)]%
        {han2023blendmr}
\bibfield{author}{\bibinfo{person}{Violet~Yinuo Han}, \bibinfo{person}{Hyunsung
  Cho}, \bibinfo{person}{Kiyosu Maeda}, \bibinfo{person}{Alexandra Ion}, {and}
  \bibinfo{person}{David Lindlbauer}.} \bibinfo{year}{2023}\natexlab{}.
\newblock \showarticletitle{BlendMR: A Computational Method to Create Ambient
  Mixed Reality Interfaces}.
\newblock \bibinfo{journal}{\emph{Proceedings of the ACM on Human-Computer
  Interaction}} \bibinfo{volume}{7}, \bibinfo{number}{ISS}
  (\bibinfo{year}{2023}), \bibinfo{pages}{217--241}.
\newblock


\bibitem[Herr et~al\mbox{.}(2018)]%
        {herr2018immersive}
\bibfield{author}{\bibinfo{person}{Dominik Herr}, \bibinfo{person}{Jan
  Reinhardt}, \bibinfo{person}{Guido Reina}, \bibinfo{person}{Robert
  Kr{\"u}ger}, \bibinfo{person}{Rafael~V Ferrari}, {and}
  \bibinfo{person}{Thomas Ertl}.} \bibinfo{year}{2018}\natexlab{}.
\newblock \showarticletitle{Immersive modular factory layout planning using
  augmented reality}.
\newblock \bibinfo{journal}{\emph{Procedia CIRP}}  \bibinfo{volume}{72}
  (\bibinfo{year}{2018}), \bibinfo{pages}{1112--1117}.
\newblock


\bibitem[Herskovitz et~al\mbox{.}(2022)]%
        {herskovitz2022xspace}
\bibfield{author}{\bibinfo{person}{Jaylin Herskovitz}, \bibinfo{person}{Yi~Fei
  Cheng}, \bibinfo{person}{Anhong Guo}, \bibinfo{person}{Alanson~P Sample},
  {and} \bibinfo{person}{Michael Nebeling}.} \bibinfo{year}{2022}\natexlab{}.
\newblock \showarticletitle{XSpace: An Augmented Reality Toolkit for Enabling
  Spatially-Aware Distributed Collaboration}.
\newblock \bibinfo{journal}{\emph{Proceedings of the ACM on Human-Computer
  Interaction}} \bibinfo{volume}{6}, \bibinfo{number}{ISS}
  (\bibinfo{year}{2022}), \bibinfo{pages}{277--302}.
\newblock


\bibitem[Huo et~al\mbox{.}(2018)]%
        {huo2018scenariot}
\bibfield{author}{\bibinfo{person}{Ke Huo}, \bibinfo{person}{Yuanzhi Cao},
  \bibinfo{person}{Sang~Ho Yoon}, \bibinfo{person}{Zhuangying Xu},
  \bibinfo{person}{Guiming Chen}, {and} \bibinfo{person}{Karthik Ramani}.}
  \bibinfo{year}{2018}\natexlab{}.
\newblock \showarticletitle{Scenariot: Spatially mapping smart things within
  augmented reality scenes}. In \bibinfo{booktitle}{\emph{Proceedings of the
  2018 CHI Conference on human factors in computing systems}}.
  \bibinfo{pages}{1--13}.
\newblock


\bibitem[Huo et~al\mbox{.}(2017)]%
        {Huo_windowshaping_2017}
\bibfield{author}{\bibinfo{person}{Ke Huo}, \bibinfo{person}{Vinayak}, {and}
  \bibinfo{person}{Karthik Ramani}.} \bibinfo{year}{2017}\natexlab{}.
\newblock \showarticletitle{Window-Shaping: 3D Design Ideation by Creating on,
  Borrowing from, and Looking at the Physical World}. In
  \bibinfo{booktitle}{\emph{Proceedings of the Eleventh International
  Conference on Tangible, Embedded, and Embodied Interaction}}.
  \bibinfo{publisher}{ACM}, \bibinfo{address}{Yokohama Japan},
  \bibinfo{pages}{37–45}.
\newblock
\showISBNx{978-1-4503-4676-4}
\urldef\tempurl%
\url{https://doi.org/10.1145/3024969.3024995}
\showDOI{\tempurl}


\bibitem[Huynh et~al\mbox{.}(2022)]%
        {huynh2022layerable}
\bibfield{author}{\bibinfo{person}{Brandon Huynh}, \bibinfo{person}{Abby
  Wysopal}, \bibinfo{person}{Vivian Ross}, \bibinfo{person}{Jason Orlosky},
  {and} \bibinfo{person}{Tobias H{\"o}llerer}.}
  \bibinfo{year}{2022}\natexlab{}.
\newblock \showarticletitle{Layerable Apps: Comparing Concurrent and Exclusive
  Display of Augmented Reality Applications}. In \bibinfo{booktitle}{\emph{2022
  IEEE International Symposium on Mixed and Augmented Reality (ISMAR)}}. IEEE,
  \bibinfo{pages}{857--863}.
\newblock


\bibitem[Jetter et~al\mbox{.}(2020)]%
        {Jetter_InVREverything_2020}
\bibfield{author}{\bibinfo{person}{Hans-Christian Jetter},
  \bibinfo{person}{Roman Rädle}, \bibinfo{person}{Tiare Feuchtner},
  \bibinfo{person}{Christoph Anthes}, \bibinfo{person}{Judith Friedl}, {and}
  \bibinfo{person}{Clemens~Nylandsted Klokmose}.}
  \bibinfo{year}{2020}\natexlab{}.
\newblock \showarticletitle{“In VR, everything is possible!”: Sketching and
  Simulating Spatially-Aware Interactive Spaces in Virtual Reality}. In
  \bibinfo{booktitle}{\emph{Proceedings of the 2020 CHI Conference on Human
  Factors in Computing Systems}}. \bibinfo{publisher}{ACM},
  \bibinfo{address}{Honolulu HI USA}, \bibinfo{pages}{1–16}.
\newblock
\showISBNx{978-1-4503-6708-0}
\urldef\tempurl%
\url{https://doi.org/10.1145/3313831.3376652}
\showDOI{\tempurl}


\bibitem[Jo and Kim(2016)]%
        {jo2016ariot}
\bibfield{author}{\bibinfo{person}{Dongsik Jo} {and}
  \bibinfo{person}{Gerard~Jounghyun Kim}.} \bibinfo{year}{2016}\natexlab{}.
\newblock \showarticletitle{ARIoT: scalable augmented reality framework for
  interacting with Internet of Things appliances everywhere}.
\newblock \bibinfo{journal}{\emph{IEEE Transactions on Consumer Electronics}}
  \bibinfo{volume}{62}, \bibinfo{number}{3} (\bibinfo{year}{2016}),
  \bibinfo{pages}{334--340}.
\newblock


\bibitem[Julier et~al\mbox{.}(2000)]%
        {julier2000information}
\bibfield{author}{\bibinfo{person}{Simon Julier}, \bibinfo{person}{Marco
  Lanzagorta}, \bibinfo{person}{Yohan Baillot}, \bibinfo{person}{Lawrence
  Rosenblum}, \bibinfo{person}{Steven Feiner}, \bibinfo{person}{Tobias
  Hollerer}, {and} \bibinfo{person}{Sabrina Sestito}.}
  \bibinfo{year}{2000}\natexlab{}.
\newblock \showarticletitle{Information filtering for mobile augmented
  reality}. In \bibinfo{booktitle}{\emph{Proceedings IEEE and ACM International
  Symposium on Augmented Reality (ISAR 2000)}}. IEEE, \bibinfo{pages}{3--11}.
\newblock


\bibitem[Kato and Billinghurst(1999)]%
        {Kato_MarkerTracking_1999}
\bibfield{author}{\bibinfo{person}{Hirokazu Kato} {and} \bibinfo{person}{Mark
  Billinghurst}.} \bibinfo{year}{1999}\natexlab{}.
\newblock \showarticletitle{Marker Tracking and {{HMD}} Calibration for a
  Video-Based Augmented Reality Conferencing System}. In
  \bibinfo{booktitle}{\emph{Proceedings 2nd {{IEEE}} and {{ACM International
  Workshop}} on {{Augmented Reality}} ({{IWAR}}'99)}}.
  \bibinfo{publisher}{{IEEE Comput. Soc}}, \bibinfo{address}{{San Francisco,
  CA, USA}}, \bibinfo{pages}{85--94}.
\newblock
\showISBNx{978-0-7695-0359-2}
\urldef\tempurl%
\url{https://doi.org/10.1109/IWAR.1999.803809}
\showDOI{\tempurl}


\bibitem[Kim et~al\mbox{.}(2019)]%
        {kim2019x}
\bibfield{author}{\bibinfo{person}{Donghwi Kim}, \bibinfo{person}{Sooyoung
  Park}, \bibinfo{person}{Jihoon Ko}, \bibinfo{person}{Steven~Y Ko}, {and}
  \bibinfo{person}{Sung-Ju Lee}.} \bibinfo{year}{2019}\natexlab{}.
\newblock \showarticletitle{X-droid: a quick and easy android prototyping
  framework with a single-app illusion}. In
  \bibinfo{booktitle}{\emph{Proceedings of the 32nd Annual ACM Symposium on
  User Interface Software and Technology}}. \bibinfo{pages}{95--108}.
\newblock


\bibitem[Kumar et~al\mbox{.}(2013)]%
        {Kumar_Webzeitgeist_2013}
\bibfield{author}{\bibinfo{person}{Ranjitha Kumar}, \bibinfo{person}{Arvind
  Satyanarayan}, \bibinfo{person}{Cesar Torres}, \bibinfo{person}{Maxine Lim},
  \bibinfo{person}{Salman Ahmad}, \bibinfo{person}{Scott~R. Klemmer}, {and}
  \bibinfo{person}{Jerry~O. Talton}.} \bibinfo{year}{2013}\natexlab{}.
\newblock \showarticletitle{Webzeitgeist: design mining the web}. In
  \bibinfo{booktitle}{\emph{Proceedings of the SIGCHI Conference on Human
  Factors in Computing Systems}}. \bibinfo{publisher}{ACM},
  \bibinfo{address}{Paris France}, \bibinfo{pages}{3083–3092}.
\newblock
\showISBNx{978-1-4503-1899-0}
\urldef\tempurl%
\url{https://doi.org/10.1145/2470654.2466420}
\showDOI{\tempurl}


\bibitem[Kumar et~al\mbox{.}(2011)]%
        {Kumar_Bricolage_2011}
\bibfield{author}{\bibinfo{person}{Ranjitha Kumar}, \bibinfo{person}{Jerry~O.
  Talton}, \bibinfo{person}{Salman Ahmad}, {and} \bibinfo{person}{Scott~R.
  Klemmer}.} \bibinfo{year}{2011}\natexlab{}.
\newblock \showarticletitle{Bricolage: example-based retargeting for web
  design}. In \bibinfo{booktitle}{\emph{Proceedings of the SIGCHI Conference on
  Human Factors in Computing Systems}}. \bibinfo{publisher}{ACM},
  \bibinfo{address}{Vancouver BC Canada}, \bibinfo{pages}{2197–2206}.
\newblock
\showISBNx{978-1-4503-0228-9}
\urldef\tempurl%
\url{https://doi.org/10.1145/1978942.1979262}
\showDOI{\tempurl}


\bibitem[Kwan and Fu(2019)]%
        {kwan2019mobi3dsketch}
\bibfield{author}{\bibinfo{person}{Kin~Chung Kwan} {and}
  \bibinfo{person}{Hongbo Fu}.} \bibinfo{year}{2019}\natexlab{}.
\newblock \showarticletitle{Mobi3dsketch: 3d sketching in mobile ar}. In
  \bibinfo{booktitle}{\emph{Proceedings of the 2019 CHI Conference on Human
  Factors in Computing Systems}}. \bibinfo{pages}{1--11}.
\newblock


\bibitem[Lages and Bowman(2019)]%
        {Lages_Bowman_2019c}
\bibfield{author}{\bibinfo{person}{Wallace~S. Lages} {and}
  \bibinfo{person}{Doug~A. Bowman}.} \bibinfo{year}{2019}\natexlab{}.
\newblock \showarticletitle{Walking with adaptive augmented reality workspaces:
  design and usage patterns}. In \bibinfo{booktitle}{\emph{Proceedings of the
  24th International Conference on Intelligent User Interfaces}}.
  \bibinfo{publisher}{ACM}, \bibinfo{address}{Marina del Ray California},
  \bibinfo{pages}{356–366}.
\newblock
\showISBNx{978-1-4503-6272-6}
\urldef\tempurl%
\url{https://doi.org/10.1145/3301275.3302278}
\showDOI{\tempurl}


\bibitem[Lang et~al\mbox{.}(2019)]%
        {lang2019virtual}
\bibfield{author}{\bibinfo{person}{Yining Lang}, \bibinfo{person}{Wei Liang},
  {and} \bibinfo{person}{Lap-Fai Yu}.} \bibinfo{year}{2019}\natexlab{}.
\newblock \showarticletitle{Virtual agent positioning driven by scene semantics
  in mixed reality}. In \bibinfo{booktitle}{\emph{2019 IEEE Conference on
  Virtual Reality and 3D User Interfaces (VR)}}. IEEE,
  \bibinfo{pages}{767--775}.
\newblock


\bibitem[Lee et~al\mbox{.}(2022a)]%
        {Lee2022}
\bibfield{author}{\bibinfo{person}{Jaewook Lee}, \bibinfo{person}{Fanjie Jin},
  \bibinfo{person}{Younsoo Kim}, {and} \bibinfo{person}{David Lindlbauer}.}
  \bibinfo{year}{2022}\natexlab{a}.
\newblock \showarticletitle{User Preference for Navigation Instructions in
  Mixed Reality}. In \bibinfo{booktitle}{\emph{2022 {IEEE} Conference on
  Virtual Reality and {3D} User Interfaces ({VR})}}. \bibinfo{pages}{802--811}.
\newblock
\showISSN{2642-5254}
\urldef\tempurl%
\url{https://doi.org/10.1109/VR51125.2022.00102}
\showDOI{\tempurl}


\bibitem[Lee et~al\mbox{.}(2022b)]%
        {Lee_Kim_Kim_Lee_Kim_Song_Ko_Oh_Shin_2022}
\bibfield{author}{\bibinfo{person}{Sunjae Lee}, \bibinfo{person}{Hoyoung Kim},
  \bibinfo{person}{Sijung Kim}, \bibinfo{person}{Sangwook Lee},
  \bibinfo{person}{Hyosu Kim}, \bibinfo{person}{Jean~Young Song},
  \bibinfo{person}{Steven~Y. Ko}, \bibinfo{person}{Sangeun Oh}, {and}
  \bibinfo{person}{Insik Shin}.} \bibinfo{year}{2022}\natexlab{b}.
\newblock \showarticletitle{A-mash: providing single-app illusion for multi-app
  use through user-centric UI mashup}. In \bibinfo{booktitle}{\emph{Proceedings
  of the 28th Annual International Conference on Mobile Computing And
  Networking}}. \bibinfo{publisher}{ACM}, \bibinfo{address}{Sydney NSW
  Australia}, \bibinfo{pages}{690–702}.
\newblock
\showISBNx{978-1-4503-9181-8}
\urldef\tempurl%
\url{https://doi.org/10.1145/3495243.3560522}
\showDOI{\tempurl}


\bibitem[Leiva et~al\mbox{.}(2021)]%
        {Leiva_RapidoPrototyping_2021}
\bibfield{author}{\bibinfo{person}{Germ{\'a}n Leiva},
  \bibinfo{person}{Jens~Emil Gr{\o}nb{\ae}k},
  \bibinfo{person}{Clemens~Nylandsted Klokmose}, \bibinfo{person}{Cuong
  Nguyen}, \bibinfo{person}{Rubaiat~Habib Kazi}, {and} \bibinfo{person}{Paul
  Asente}.} \bibinfo{year}{2021}\natexlab{}.
\newblock \showarticletitle{Rapido: {{Prototyping Interactive AR Experiences}}
  through {{Programming}} by {{Demonstration}}}. In
  \bibinfo{booktitle}{\emph{The 34th {{Annual ACM Symposium}} on {{User
  Interface Software}} and {{Technology}}}}. \bibinfo{publisher}{{ACM}},
  \bibinfo{address}{{Virtual Event USA}}, \bibinfo{pages}{626--637}.
\newblock
\showISBNx{978-1-4503-8635-7}
\urldef\tempurl%
\url{https://doi.org/10.1145/3472749.3474774}
\showDOI{\tempurl}


\bibitem[Leiva et~al\mbox{.}(2019)]%
        {Leiva_EnactReducing_2019}
\bibfield{author}{\bibinfo{person}{Germ{\'a}n Leiva}, \bibinfo{person}{Nolwenn
  Maudet}, \bibinfo{person}{Wendy Mackay}, {and} \bibinfo{person}{Michel
  {Beaudouin-Lafon}}.} \bibinfo{year}{2019}\natexlab{}.
\newblock \showarticletitle{Enact: {{Reducing Designer}}\textendash{{Developer
  Breakdowns When Prototyping Custom Interactions}}}.
\newblock \bibinfo{journal}{\emph{ACM Transactions on Computer-Human
  Interaction}} \bibinfo{volume}{26}, \bibinfo{number}{3} (\bibinfo{date}{June}
  \bibinfo{year}{2019}), \bibinfo{pages}{1--48}.
\newblock
\showISSN{1073-0516, 1557-7325}
\urldef\tempurl%
\url{https://doi.org/10.1145/3310276}
\showDOI{\tempurl}


\bibitem[Leiva et~al\mbox{.}(2020)]%
        {Leiva_ProntoRapid_2020a}
\bibfield{author}{\bibinfo{person}{Germ{\'a}n Leiva}, \bibinfo{person}{Cuong
  Nguyen}, \bibinfo{person}{Rubaiat~Habib Kazi}, {and} \bibinfo{person}{Paul
  Asente}.} \bibinfo{year}{2020}\natexlab{}.
\newblock \showarticletitle{Pronto: {{Rapid Augmented Reality Video Prototyping
  Using Sketches}} and {{Enaction}}}. In \bibinfo{booktitle}{\emph{Proceedings
  of the 2020 {{CHI Conference}} on {{Human Factors}} in {{Computing
  Systems}}}}. \bibinfo{publisher}{{ACM}}, \bibinfo{address}{{Honolulu HI
  USA}}, \bibinfo{pages}{1--13}.
\newblock
\showISBNx{978-1-4503-6708-0}
\urldef\tempurl%
\url{https://doi.org/10.1145/3313831.3376160}
\showDOI{\tempurl}


\bibitem[Li et~al\mbox{.}(2017)]%
        {Li_Azaria_Myers_2017}
\bibfield{author}{\bibinfo{person}{Toby Jia-Jun Li}, \bibinfo{person}{Amos
  Azaria}, {and} \bibinfo{person}{Brad~A. Myers}.}
  \bibinfo{year}{2017}\natexlab{}.
\newblock \showarticletitle{SUGILITE: Creating Multimodal Smartphone Automation
  by Demonstration}. In \bibinfo{booktitle}{\emph{Proceedings of the 2017 CHI
  Conference on Human Factors in Computing Systems}}. \bibinfo{publisher}{ACM},
  \bibinfo{address}{Denver Colorado USA}, \bibinfo{pages}{6038–6049}.
\newblock
\showISBNx{978-1-4503-4655-9}
\urldef\tempurl%
\url{https://doi.org/10.1145/3025453.3025483}
\showDOI{\tempurl}


\bibitem[Li et~al\mbox{.}(2021)]%
        {Li_Popowski_Mitchell_Myers_2021}
\bibfield{author}{\bibinfo{person}{Toby Jia-Jun Li}, \bibinfo{person}{Lindsay
  Popowski}, \bibinfo{person}{Tom Mitchell}, {and} \bibinfo{person}{Brad~A
  Myers}.} \bibinfo{year}{2021}\natexlab{}.
\newblock \showarticletitle{Screen2Vec: Semantic Embedding of GUI Screens and
  GUI Components}. In \bibinfo{booktitle}{\emph{Proceedings of the 2021 CHI
  Conference on Human Factors in Computing Systems}}. \bibinfo{publisher}{ACM},
  \bibinfo{address}{Yokohama Japan}, \bibinfo{pages}{1–15}.
\newblock
\showISBNx{978-1-4503-8096-6}
\urldef\tempurl%
\url{https://doi.org/10.1145/3411764.3445049}
\showDOI{\tempurl}


\bibitem[Li and Riva(2018)]%
        {Li_Riva_2018}
\bibfield{author}{\bibinfo{person}{Toby Jia-Jun Li} {and}
  \bibinfo{person}{Oriana Riva}.} \bibinfo{year}{2018}\natexlab{}.
\newblock \showarticletitle{Kite: Building Conversational Bots from Mobile
  Apps}. In \bibinfo{booktitle}{\emph{Proceedings of the 16th Annual
  International Conference on Mobile Systems, Applications, and Services}}.
  \bibinfo{publisher}{ACM}, \bibinfo{address}{Munich Germany},
  \bibinfo{pages}{96–109}.
\newblock
\showISBNx{978-1-4503-5720-3}
\urldef\tempurl%
\url{https://doi.org/10.1145/3210240.3210339}
\showDOI{\tempurl}


\bibitem[Liang et~al\mbox{.}(2021)]%
        {liang2021scene}
\bibfield{author}{\bibinfo{person}{Wei Liang}, \bibinfo{person}{Xinzhe Yu},
  \bibinfo{person}{Rawan Alghofaili}, \bibinfo{person}{Yining Lang}, {and}
  \bibinfo{person}{Lap-Fai Yu}.} \bibinfo{year}{2021}\natexlab{}.
\newblock \showarticletitle{Scene-aware behavior synthesis for virtual pets in
  mixed reality}. In \bibinfo{booktitle}{\emph{Proceedings of the 2021 CHI
  Conference on Human Factors in Computing Systems}}. \bibinfo{pages}{1--12}.
\newblock


\bibitem[Lindlbauer et~al\mbox{.}(2019)]%
        {Lindlbauer_ContextAwareMR_2019}
\bibfield{author}{\bibinfo{person}{David Lindlbauer},
  \bibinfo{person}{Anna~Maria Feit}, {and} \bibinfo{person}{Otmar Hilliges}.}
  \bibinfo{year}{2019}\natexlab{}.
\newblock \showarticletitle{Context-Aware Online Adaptation of Mixed Reality
  Interfaces}. In \bibinfo{booktitle}{\emph{Proceedings of the 32nd Annual ACM
  Symposium on User Interface Software and Technology}}.
  \bibinfo{publisher}{ACM}, \bibinfo{address}{New Orleans LA USA},
  \bibinfo{pages}{147–160}.
\newblock
\showISBNx{978-1-4503-6816-2}
\urldef\tempurl%
\url{https://doi.org/10.1145/3332165.3347945}
\showDOI{\tempurl}


\bibitem[Lu and Bowman(2021)]%
        {Lu_Bowman_2021}
\bibfield{author}{\bibinfo{person}{Feiyu Lu} {and} \bibinfo{person}{Doug~A.
  Bowman}.} \bibinfo{year}{2021}\natexlab{}.
\newblock \showarticletitle{Evaluating the Potential of Glanceable AR
  Interfaces for Authentic Everyday Uses}. In \bibinfo{booktitle}{\emph{2021
  IEEE Virtual Reality and 3D User Interfaces (VR)}}.
  \bibinfo{publisher}{IEEE}, \bibinfo{address}{Lisboa, Portugal},
  \bibinfo{pages}{768–777}.
\newblock
\showISBNx{978-1-66541-838-6}
\urldef\tempurl%
\url{https://doi.org/10.1109/VR50410.2021.00104}
\showDOI{\tempurl}


\bibitem[Lu et~al\mbox{.}(2020)]%
        {Lu_Davari_Lisle_Li_Bowman_2020}
\bibfield{author}{\bibinfo{person}{Feiyu Lu}, \bibinfo{person}{Shakiba Davari},
  \bibinfo{person}{Lee Lisle}, \bibinfo{person}{Yuan Li}, {and}
  \bibinfo{person}{Doug~A. Bowman}.} \bibinfo{year}{2020}\natexlab{}.
\newblock \showarticletitle{Glanceable AR: Evaluating Information Access
  Methods for Head-Worn Augmented Reality}. In \bibinfo{booktitle}{\emph{2020
  IEEE Conference on Virtual Reality and 3D User Interfaces (VR)}}.
  \bibinfo{publisher}{IEEE}, \bibinfo{address}{Atlanta, GA, USA},
  \bibinfo{pages}{930–939}.
\newblock
\showISBNx{978-1-72815-608-8}
\urldef\tempurl%
\url{https://doi.org/10.1109/VR46266.2020.00113}
\showDOI{\tempurl}


\bibitem[Lu and Xu(2022)]%
        {Lu_Xu_2022}
\bibfield{author}{\bibinfo{person}{Feiyu Lu} {and} \bibinfo{person}{Yan Xu}.}
  \bibinfo{year}{2022}\natexlab{}.
\newblock \showarticletitle{Exploring Spatial UI Transition Mechanisms with
  Head-Worn Augmented Reality}. In \bibinfo{booktitle}{\emph{CHI Conference on
  Human Factors in Computing Systems}}. \bibinfo{publisher}{ACM},
  \bibinfo{address}{New Orleans LA USA}, \bibinfo{pages}{1–16}.
\newblock
\showISBNx{978-1-4503-9157-3}
\urldef\tempurl%
\url{https://doi.org/10.1145/3491102.3517723}
\showDOI{\tempurl}


\bibitem[MacIntyre et~al\mbox{.}(2004)]%
        {macintyre2004dart}
\bibfield{author}{\bibinfo{person}{Blair MacIntyre}, \bibinfo{person}{Maribeth
  Gandy}, \bibinfo{person}{Steven Dow}, {and} \bibinfo{person}{Jay~David
  Bolter}.} \bibinfo{year}{2004}\natexlab{}.
\newblock \showarticletitle{DART: a toolkit for rapid design exploration of
  augmented reality experiences}. In \bibinfo{booktitle}{\emph{Proceedings of
  the 17th annual ACM symposium on User interface software and technology}}.
  \bibinfo{pages}{197--206}.
\newblock


\bibitem[Mine(1995)]%
        {Mine_ISAAC_1995}
\bibfield{author}{\bibinfo{person}{Mark~R Mine}.}
  \bibinfo{year}{1995}\natexlab{}.
\newblock \showarticletitle{ISAAC: A Virtual Environment Tool for the
  Interactive Construction of Virtual Worlds}.
\newblock  (\bibinfo{year}{1995}).
\newblock


\bibitem[Nebeling et~al\mbox{.}(2018)]%
        {Nebeling_ProtoARRapid_2018}
\bibfield{author}{\bibinfo{person}{Michael Nebeling}, \bibinfo{person}{Janet
  Nebeling}, \bibinfo{person}{Ao Yu}, {and} \bibinfo{person}{Rob Rumble}.}
  \bibinfo{year}{2018}\natexlab{}.
\newblock \showarticletitle{{{ProtoAR}}: {{Rapid Physical-Digital Prototyping}}
  of {{Mobile Augmented Reality Applications}}}. In
  \bibinfo{booktitle}{\emph{Proceedings of the 2018 {{CHI Conference}} on
  {{Human Factors}} in {{Computing Systems}}}}. \bibinfo{publisher}{{ACM}},
  \bibinfo{address}{{Montreal QC Canada}}, \bibinfo{pages}{1--12}.
\newblock
\showISBNx{978-1-4503-5620-6}
\urldef\tempurl%
\url{https://doi.org/10.1145/3173574.3173927}
\showDOI{\tempurl}


\bibitem[Nebeling and Speicher(2018)]%
        {Nebeling2018_trouble}
\bibfield{author}{\bibinfo{person}{Michael Nebeling} {and}
  \bibinfo{person}{Maximilian Speicher}.} \bibinfo{year}{2018}\natexlab{}.
\newblock \showarticletitle{The trouble with augmented reality/virtual reality
  authoring tools}. In \bibinfo{booktitle}{\emph{2018 IEEE International
  Symposium on Mixed and Augmented Reality Adjunct (ISMAR-Adjunct)}}.
  \bibinfo{publisher}{IEEE}.
\newblock
\showISBNx{9781538675922}
\urldef\tempurl%
\url{https://doi.org/10.1109/ismar-adjunct.2018.00098}
\showDOI{\tempurl}


\bibitem[Nebeling et~al\mbox{.}(2020)]%
        {nebeling2020mrat}
\bibfield{author}{\bibinfo{person}{Michael Nebeling},
  \bibinfo{person}{Maximilian Speicher}, \bibinfo{person}{Xizi Wang},
  \bibinfo{person}{Shwetha Rajaram}, \bibinfo{person}{Brian~D Hall},
  \bibinfo{person}{Zijian Xie}, \bibinfo{person}{Alexander~RE Raistrick},
  \bibinfo{person}{Michelle Aebersold}, \bibinfo{person}{Edward~G Happ},
  \bibinfo{person}{Jiayin Wang}, {et~al\mbox{.}}}
  \bibinfo{year}{2020}\natexlab{}.
\newblock \showarticletitle{MRAT: The mixed reality analytics toolkit}. In
  \bibinfo{booktitle}{\emph{Proceedings of the 2020 CHI Conference on Human
  Factors in Computing Systems}}. \bibinfo{pages}{1--12}.
\newblock


\bibitem[Nuernberger et~al\mbox{.}(2016)]%
        {Nuernberger_Ofek_Benko_Wilson_2016}
\bibfield{author}{\bibinfo{person}{Benjamin Nuernberger}, \bibinfo{person}{Eyal
  Ofek}, \bibinfo{person}{Hrvoje Benko}, {and} \bibinfo{person}{Andrew~D.
  Wilson}.} \bibinfo{year}{2016}\natexlab{}.
\newblock \showarticletitle{SnapToReality: Aligning Augmented Reality to the
  Real World}. In \bibinfo{booktitle}{\emph{Proceedings of the 2016 CHI
  Conference on Human Factors in Computing Systems}}. \bibinfo{publisher}{ACM},
  \bibinfo{address}{San Jose California USA}, \bibinfo{pages}{1233–1244}.
\newblock
\showISBNx{978-1-4503-3362-7}
\urldef\tempurl%
\url{https://doi.org/10.1145/2858036.2858250}
\showDOI{\tempurl}


\bibitem[Prouzeau et~al\mbox{.}(2020)]%
        {Prouzeau_CorsicanTwin_2020}
\bibfield{author}{\bibinfo{person}{Arnaud Prouzeau}, \bibinfo{person}{Yuchen
  Wang}, \bibinfo{person}{Barrett Ens}, \bibinfo{person}{Wesley Willett}, {and}
  \bibinfo{person}{Tim Dwyer}.} \bibinfo{year}{2020}\natexlab{}.
\newblock \showarticletitle{Corsican Twin: Authoring In Situ Augmented Reality
  Visualisations in Virtual Reality}. In \bibinfo{booktitle}{\emph{Proceedings
  of the International Conference on Advanced Visual Interfaces}}.
  \bibinfo{publisher}{ACM}, \bibinfo{address}{Salerno Italy},
  \bibinfo{pages}{1–9}.
\newblock
\showISBNx{978-1-4503-7535-1}
\urldef\tempurl%
\url{https://doi.org/10.1145/3399715.3399743}
\showDOI{\tempurl}


\bibitem[Qian et~al\mbox{.}(2022)]%
        {Qian_He_Hu_Wang_Ipsita_Ramani_2022}
\bibfield{author}{\bibinfo{person}{Xun Qian}, \bibinfo{person}{Fengming He},
  \bibinfo{person}{Xiyun Hu}, \bibinfo{person}{Tianyi Wang},
  \bibinfo{person}{Ananya Ipsita}, {and} \bibinfo{person}{Karthik Ramani}.}
  \bibinfo{year}{2022}\natexlab{}.
\newblock \showarticletitle{ScalAR: Authoring Semantically Adaptive Augmented
  Reality Experiences in Virtual Reality}. In \bibinfo{booktitle}{\emph{CHI
  Conference on Human Factors in Computing Systems}}. \bibinfo{publisher}{ACM},
  \bibinfo{address}{New Orleans LA USA}, \bibinfo{pages}{1–18}.
\newblock
\showISBNx{978-1-4503-9157-3}
\urldef\tempurl%
\url{https://doi.org/10.1145/3491102.3517665}
\showDOI{\tempurl}


\bibitem[Romero-Ramirez et~al\mbox{.}(2018)]%
        {romero2018speeded}
\bibfield{author}{\bibinfo{person}{Francisco~J Romero-Ramirez},
  \bibinfo{person}{Rafael Mu{\~n}oz-Salinas}, {and} \bibinfo{person}{Rafael
  Medina-Carnicer}.} \bibinfo{year}{2018}\natexlab{}.
\newblock \showarticletitle{Speeded up detection of squared fiducial markers}.
\newblock \bibinfo{journal}{\emph{Image and vision Computing}}
  \bibinfo{volume}{76} (\bibinfo{year}{2018}), \bibinfo{pages}{38--47}.
\newblock


\bibitem[Seichter et~al\mbox{.}(2008)]%
        {Seichter_Looser_Billinghurst_2008}
\bibfield{author}{\bibinfo{person}{Hartmut Seichter}, \bibinfo{person}{Julian
  Looser}, {and} \bibinfo{person}{Mark Billinghurst}.}
  \bibinfo{year}{2008}\natexlab{}.
\newblock \showarticletitle{ComposAR: An intuitive tool for authoring AR
  applications}. In \bibinfo{booktitle}{\emph{2008 7th IEEE/ACM International
  Symposium on Mixed and Augmented Reality}}. \bibinfo{publisher}{IEEE},
  \bibinfo{address}{Cambridge, UK}, \bibinfo{pages}{177–178}.
\newblock
\showISBNx{978-1-4244-2840-3}
\urldef\tempurl%
\url{https://doi.org/10.1109/ISMAR.2008.4637354}
\showDOI{\tempurl}


\bibitem[Spatial(2023)]%
        {spatial}
\bibfield{author}{\bibinfo{person}{Spatial}.} \bibinfo{year}{2023}\natexlab{}.
\newblock \bibinfo{title}{Create, share, and experience your creativity in 3D}.
\newblock
\newblock
\urldef\tempurl%
\url{http://www.spatial.io/}
\showURL{%
\tempurl}


\bibitem[Tahara et~al\mbox{.}(2020)]%
        {tahara2020retargetable}
\bibfield{author}{\bibinfo{person}{Tomu Tahara}, \bibinfo{person}{Takashi
  Seno}, \bibinfo{person}{Gaku Narita}, {and} \bibinfo{person}{Tomoya
  Ishikawa}.} \bibinfo{year}{2020}\natexlab{}.
\newblock \showarticletitle{Retargetable AR: Context-aware augmented reality in
  indoor scenes based on 3D scene graph}. In \bibinfo{booktitle}{\emph{2020
  IEEE International Symposium on Mixed and Augmented Reality Adjunct
  (ISMAR-Adjunct)}}. IEEE, \bibinfo{pages}{249--255}.
\newblock


\bibitem[Tatzgern et~al\mbox{.}(2016)]%
        {tatzgern2016adaptive}
\bibfield{author}{\bibinfo{person}{Markus Tatzgern}, \bibinfo{person}{Valeria
  Orso}, \bibinfo{person}{Denis Kalkofen}, \bibinfo{person}{Giulio Jacucci},
  \bibinfo{person}{Luciano Gamberini}, {and} \bibinfo{person}{Dieter
  Schmalstieg}.} \bibinfo{year}{2016}\natexlab{}.
\newblock \showarticletitle{Adaptive information density for augmented reality
  displays}. In \bibinfo{booktitle}{\emph{2016 IEEE Virtual Reality (VR)}}.
  IEEE, \bibinfo{pages}{83--92}.
\newblock


\bibitem[Wagner and Schmalstieg(2007)]%
        {wagner2007artoolkitplus}
\bibfield{author}{\bibinfo{person}{Daniel Wagner} {and} \bibinfo{person}{Dieter
  Schmalstieg}.} \bibinfo{year}{2007}\natexlab{}.
\newblock \showarticletitle{Artoolkitplus for pose tracking on mobile devices}.
\newblock  (\bibinfo{year}{2007}).
\newblock


\bibitem[Wang et~al\mbox{.}(2023)]%
        {Wang_EnablingConversational_2023}
\bibfield{author}{\bibinfo{person}{Bryan Wang}, \bibinfo{person}{Gang Li},
  {and} \bibinfo{person}{Yang Li}.} \bibinfo{year}{2023}\natexlab{}.
\newblock \bibinfo{title}{Enabling {{Conversational Interaction}} with {{Mobile
  UI}} Using {{Large Language Models}}}.
\newblock
\newblock
\showeprint[arxiv]{arXiv:2209.08655}


\bibitem[Wang et~al\mbox{.}(2021a)]%
        {Wang_Screen2WordsAutomatic_2021}
\bibfield{author}{\bibinfo{person}{Bryan Wang}, \bibinfo{person}{Gang Li},
  \bibinfo{person}{Xin Zhou}, \bibinfo{person}{Zhourong Chen},
  \bibinfo{person}{Tovi Grossman}, {and} \bibinfo{person}{Yang Li}.}
  \bibinfo{year}{2021}\natexlab{a}.
\newblock \showarticletitle{{{Screen2Words}}: {{Automatic Mobile UI
  Summarization}} with {{Multimodal Learning}}}. In
  \bibinfo{booktitle}{\emph{The 34th {{Annual ACM Symposium}} on {{User
  Interface Software}} and {{Technology}}}}. \bibinfo{publisher}{{ACM}},
  \bibinfo{address}{{Virtual Event USA}}, \bibinfo{pages}{498--510}.
\newblock
\showISBNx{978-1-4503-8635-7}
\urldef\tempurl%
\url{https://doi.org/10.1145/3472749.3474765}
\showDOI{\tempurl}


\bibitem[Wang et~al\mbox{.}(2020)]%
        {Wang_capturar_2020}
\bibfield{author}{\bibinfo{person}{Tianyi Wang}, \bibinfo{person}{Xun Qian},
  \bibinfo{person}{Fengming He}, \bibinfo{person}{Xiyun Hu},
  \bibinfo{person}{Ke Huo}, \bibinfo{person}{Yuanzhi Cao}, {and}
  \bibinfo{person}{Karthik Ramani}.} \bibinfo{year}{2020}\natexlab{}.
\newblock \showarticletitle{CAPturAR: An Augmented Reality Tool for Authoring
  Human-Involved Context-Aware Applications}. In
  \bibinfo{booktitle}{\emph{Proceedings of the 33rd Annual ACM Symposium on
  User Interface Software and Technology}}. \bibinfo{publisher}{ACM},
  \bibinfo{address}{Virtual Event USA}, \bibinfo{pages}{328–341}.
\newblock
\showISBNx{978-1-4503-7514-6}
\urldef\tempurl%
\url{https://doi.org/10.1145/3379337.3415815}
\showDOI{\tempurl}


\bibitem[Wang et~al\mbox{.}(2021b)]%
        {Wang_DistanciAR_2021}
\bibfield{author}{\bibinfo{person}{Zeyu Wang}, \bibinfo{person}{Cuong Nguyen},
  \bibinfo{person}{Paul Asente}, {and} \bibinfo{person}{Julie Dorsey}.}
  \bibinfo{year}{2021}\natexlab{b}.
\newblock \showarticletitle{DistanciAR: Authoring Site-Specific Augmented
  Reality Experiences for Remote Environments}. In
  \bibinfo{booktitle}{\emph{Proceedings of the 2021 CHI Conference on Human
  Factors in Computing Systems}}. \bibinfo{publisher}{ACM},
  \bibinfo{address}{Yokohama Japan}, \bibinfo{pages}{1–12}.
\newblock
\showISBNx{978-1-4503-8096-6}
\urldef\tempurl%
\url{https://doi.org/10.1145/3411764.3445552}
\showDOI{\tempurl}


\bibitem[Wu et~al\mbox{.}(2021)]%
        {Wu_Zhang_Nichols_Bigham_2021}
\bibfield{author}{\bibinfo{person}{Jason Wu}, \bibinfo{person}{Xiaoyi Zhang},
  \bibinfo{person}{Jeff Nichols}, {and} \bibinfo{person}{Jeffrey~P Bigham}.}
  \bibinfo{year}{2021}\natexlab{}.
\newblock \showarticletitle{Screen Parsing: Towards Reverse Engineering of UI
  Models from Screenshots}. In \bibinfo{booktitle}{\emph{The 34th Annual ACM
  Symposium on User Interface Software and Technology}}.
  \bibinfo{publisher}{ACM}, \bibinfo{address}{Virtual Event USA},
  \bibinfo{pages}{470–483}.
\newblock
\showISBNx{978-1-4503-8635-7}
\urldef\tempurl%
\url{https://doi.org/10.1145/3472749.3474763}
\showDOI{\tempurl}


\bibitem[Xia et~al\mbox{.}(2018)]%
        {Xia_Spacetime_2018}
\bibfield{author}{\bibinfo{person}{Haijun Xia}, \bibinfo{person}{Sebastian
  Herscher}, \bibinfo{person}{Ken Perlin}, {and} \bibinfo{person}{Daniel
  Wigdor}.} \bibinfo{year}{2018}\natexlab{}.
\newblock \showarticletitle{Spacetime: Enabling Fluid Individual and
  Collaborative Editing in Virtual Reality}. In
  \bibinfo{booktitle}{\emph{Proceedings of the 31st Annual ACM Symposium on
  User Interface Software and Technology}}. \bibinfo{publisher}{ACM},
  \bibinfo{address}{Berlin Germany}, \bibinfo{pages}{853–866}.
\newblock
\showISBNx{978-1-4503-5948-1}
\urldef\tempurl%
\url{https://doi.org/10.1145/3242587.3242597}
\showDOI{\tempurl}


\bibitem[Yue et~al\mbox{.}(2017)]%
        {Yue_scenectrl_2017}
\bibfield{author}{\bibinfo{person}{Ya-Ting Yue}, \bibinfo{person}{Yong-Liang
  Yang}, \bibinfo{person}{Gang Ren}, {and} \bibinfo{person}{Wenping Wang}.}
  \bibinfo{year}{2017}\natexlab{}.
\newblock \showarticletitle{SceneCtrl: Mixed Reality Enhancement via Efficient
  Scene Editing}. In \bibinfo{booktitle}{\emph{Proceedings of the 30th Annual
  ACM Symposium on User Interface Software and Technology}}.
  \bibinfo{publisher}{ACM}, \bibinfo{address}{Québec City QC Canada},
  \bibinfo{pages}{427–436}.
\newblock
\showISBNx{978-1-4503-4981-9}
\urldef\tempurl%
\url{https://doi.org/10.1145/3126594.3126601}
\showDOI{\tempurl}


\bibitem[Zhang and Oney(2020)]%
        {Zhang_FlowMatic_2020}
\bibfield{author}{\bibinfo{person}{Lei Zhang} {and} \bibinfo{person}{Steve
  Oney}.} \bibinfo{year}{2020}\natexlab{}.
\newblock \showarticletitle{FlowMatic: An Immersive Authoring Tool for Creating
  Interactive Scenes in Virtual Reality}. In
  \bibinfo{booktitle}{\emph{Proceedings of the 33rd Annual ACM Symposium on
  User Interface Software and Technology}}. \bibinfo{publisher}{ACM},
  \bibinfo{address}{Virtual Event USA}, \bibinfo{pages}{342–353}.
\newblock
\showISBNx{978-1-4503-7514-6}
\urldef\tempurl%
\url{https://doi.org/10.1145/3379337.3415824}
\showDOI{\tempurl}


\bibitem[Zhang et~al\mbox{.}(2021)]%
        {zhang2021screen}
\bibfield{author}{\bibinfo{person}{Xiaoyi Zhang}, \bibinfo{person}{Lilian de
  Greef}, \bibinfo{person}{Amanda Swearngin}, \bibinfo{person}{Samuel White},
  \bibinfo{person}{Kyle Murray}, \bibinfo{person}{Lisa Yu}, \bibinfo{person}{Qi
  Shan}, \bibinfo{person}{Jeffrey Nichols}, \bibinfo{person}{Jason Wu},
  \bibinfo{person}{Chris Fleizach}, {et~al\mbox{.}}}
  \bibinfo{year}{2021}\natexlab{}.
\newblock \showarticletitle{Screen recognition: Creating accessibility metadata
  for mobile applications from pixels}. In
  \bibinfo{booktitle}{\emph{Proceedings of the 2021 CHI Conference on Human
  Factors in Computing Systems}}. \bibinfo{pages}{1--15}.
\newblock


\end{thebibliography}


\appendix
\setcounter{section}{0}
\vspace{1cm}
\clearpage

\renewcommand{\thesection}{\Alph{section}}
\section{Appendix}
\vspace{0.2cm}




\noindent\begin{minipage}{\fullwidthfigure}
\subsection{Top 10 functionalities by environment} 
\newcolumntype{P}[1]{>{\centering\arraybackslash}p{#1}}
\vspace{2em}
\label{sec:top_functionalities}
    \captionof{table}{Top 10 functionalities by environment}
    \hspace{0.7cm}
    \begin{tabular}{p{6cm}P{1cm}}
        \multicolumn{2}{c}{\textbf{Living Room}} \\
        \midrule
        (App/Group) Functionality & Count \\
        \midrule
        (Group) App icon & 52 \\
        (Group) UI navigation & 24 \\
        (Group) Video & 18 \\
        (Social Media App) Stories & 17 \\
        (Group) Music play/pause control & 17 \\
        (Group) Current weather information & 16 \\
        (Group) Social media post & 16 \\
        (Video Sharing App) Search & 14 \\
        (Group) Search engine search & 11 \\
        (Group) Music playlist & 9 \\
        \bottomrule
    \end{tabular}
    \vspace{1em}
    \hspace{0.5cm}
    \begin{tabular}{p{6cm}P{1cm}}
        \multicolumn{2}{c}{\textbf{Office}} \\
        \midrule
        (App/Group) Functionality & Count \\
        \midrule
        (Group) App icon & 25 \\
        (Group) Message conversation list & 19 \\
        (Group) UI navigation & 19 \\
        (Files) Reference figure & 18 \\
        (Group) Social media post & 12 \\
        (Group) Email inbox & 7 \\
        (Video Conferencing App) Video call screen & 7 \\
        (Group) Music play/pause control & 7 \\
        (Group) Video & 7 \\
        (Group) Song title and artist & 7 \\
        \bottomrule
    \end{tabular}
    \vspace{1em}
    \\
    
    \hspace{0.7cm}
    \begin{tabular}{p{6cm}P{1cm}}
        \multicolumn{2}{c}{\textbf{Coffee Shop}} \\
        \midrule
        (App/Group) Functionality & Count \\
        \midrule
        (Group) UI navigation & 45 \\
        (Group) Message conversation list & 20 \\
        (Group) Social media post & 19 \\
        (Group) App icon & 18 \\
        (Group) Music play/pause control & 17 \\
        (Group) Email inbox & 10 \\
        (Group) Search engine search & 10 \\
        (Group) To-do list & 9 \\
        (Group) Daily schedule & 9 \\
        (Social Media App) Post caption \& comments & 8 \\
        \bottomrule
    \end{tabular}
    \vspace{1em}
    \hspace{0.5cm}
    \begin{tabular}{p{6cm}P{1cm}}
        \multicolumn{2}{c}{\textbf{Kitchen}} \\
        \midrule
        (App/Group) Functionality & Count \\
        \midrule
        (Group) Instructions & 15 \\
        (Group) App icon & 14 \\
        (Group) Video & 13 \\
        (Group) Music playlist & 7 \\
        (Learning App) Language learning practice & 6 \\
        (Group) Email inbox & 5 \\
        (Group) To-do list & 4 \\
        (Group) Music play/pause control & 4 \\
        (Online Shopping App) product image & 4 \\
        (Group) Daily schedule & 4 \\
        \bottomrule
    \end{tabular}
    \vspace{1em}
    \\
    
    \hspace{0.7cm}
    \begin{tabular}{p{6cm}P{1cm}}
        \multicolumn{2}{c}{\textbf{All environments}} \\
        \midrule
        (App/Group) Functionality & Count \\
        \midrule
        (Group) App icon & 110 \\
        (Group) UI navigation & 92 \\
        (Group) Message conversation list & 46 \\
        (Group) Video & 45 \\
        (Group) Music play/pause control & 45 \\
        (Group) Music playlist & 26 \\
        (Social Media App) Stories & 26 \\
        (Group) Email inbox & 25 \\
        (Group) Current weather information & 23 \\
        (Group) Today's schedule & 21 \\
        \bottomrule
    \end{tabular}
    \vspace{1em}
\end{minipage}
\end{document}